\documentclass[twocolumn]{autart}    
\usepackage{amsmath,amssymb,bbm,color,psfrag,amsfonts,mathtools,amsbsy}
\usepackage{dsfont}
\usepackage{graphicx,cuted}                       

\usepackage{wrapfig}
\usepackage{pdfpages}
\usepackage{accents}
\usepackage{lipsum}
\usepackage[T1]{fontenc}
\usepackage[utf8]{inputenc}
\usepackage{mathtools}
\usepackage{multicol}
\usepackage{paralist}   
\usepackage{figlatex}
\let\labelindent\relax
\usepackage{enumitem}
\usepackage{booktabs}  
\usepackage{multirow}
\usepackage[normalem]{ulem}

\usepackage{caption}
\usepackage{subcaption}

\usepackage{cite}
\usepackage[pdftex,pdfpagelabels,bookmarks,hyperindex,hyperfigures]{hyperref}

\newcounter{problem}

\makeatletter
\newcommand\footnoteref[1]{\protected@xdef\@thefnmark{\ref{#1}}\@footnotemark}
\makeatother

\newtheorem{theorem}{Theorem}

\newtheorem{lemma}{Lemma}
\newtheorem{proposition}{Proposition}

\newtheorem{assumption}{Assumption}

\newtheorem{remark}{Remark}
\newtheorem{example}{Example}

\usepackage{color}

\newcommand{\kibitz}[2]{\ifnum\Comments=0\textcolor{#1}{#2}\fi}

\newcommand{\ubar}[1]{\underaccent{\bar}{#1}}

\newcommand{\real}{\mathbb{R}}

\newcommand{\mc}{\mathcal}

\newcommand{\zl}[1]{{\color{black} #1}}

\newcommand{\xunc}{x^{\text{unc}}}

\newcommand{\xup}{\hat{\bar{x}}}
\newcommand{\xlb}{\hat{\ubar{x}}}

\newcommand{\xcrit}{x^{\text{crit}}}

\newcommand{\xter}{\bar{x}^f}

\newcommand{\xlift}{\pmb{x}}

\newcommand{\lambdalift}{\pmb{\lambda}}
\newcommand{\thetalift}{\pmb{\theta}}
\newcommand{\Flift}{\pmb{F}}

\newcommand{\Ulift}{\pmb{U}}

\newcommand{\Xflift}{\pmb{\mc X_f}}
\newcommand{\Xlift}{\pmb{\mc X}}

\newcommand{\thetaup}{\bar{\theta}}
\newcommand{\thetalb}{\ubar{\theta}}
\newcommand{\lambdaup}{\bar{\lambda}}
\newcommand{\lambdalb}{\ubar{\lambda}}

\newcommand{\xjam}{x^{\text{jam}}}
\newcommand{\xjamT}{x^{\text{jam}T}}
\newcommand{\feq}{f^{\text{eq}}}

\newcommand{\fin}{f^{\text{in}}}
\newcommand{\Fin}{F^{\text{in}}}

\newcommand{\fout}{f^{\text{out}}}

\newcommand{\Fout}{F^{\text{out}}}

\newcommand{\foutm}{f^{\text{out,m}}}

\newcommand{\IR}{\mathbf{I}-\mathbf{R}}

\newcommand{\Cm}{C^{\max}}
\newcommand{\CmT}{C^{\max T}}

\edef\endfrontmatter{%
  \unexpanded\expandafter{\endfrontmatter}
  \noexpand\endNoHyper 
}

\begin{document}
	
	\begin{frontmatter}
		
		\title{Output-feedback adaptive model predictive control for ramp metering: a set-membership approach} 
		
		\thanks[footnoteinfo]{ This work was supported in part by METRANS 20-17. Z. Li also acknowledges the support from Teh-Fu Yen and Dr. Shia-Ping Siao Yen Fellowship. K. Savla has a financial interest in Xtelligent, Inc.}
		
		\author[CE]{Zhexian Li}\ead{zhexianl@usc.edu},    
		\author[CE,ISE,EE]{Ketan Savla}\ead{ksavla@usc.edu}               
		
		\address[CE]{Sonny Astani Department of Civil and Environmental Engineering, University of Southern California, Los Angeles, CA}  
		\vspace{-0.1in}
		\address[ISE]{Daniel J. Epstein Department of Industrial and Systems Engineering, University of Southern California, Los Angeles, CA}  
		\vspace{-0.1in}
		\address[EE]{Ming Hsieh Department of Electrical Engineering, University of Southern California, Los Angeles, CA}  

\begin{abstract}
Ramp metering, which regulates the flow entering the freeway, is one of the most effective freeway traffic control methods. This paper introduces an output-feedback adaptive approach to ramp metering that combines model predictive control (MPC) with set-membership parameter and state estimation. The set-membership estimator is based on a mixed-monotone embedding of underlying traffic dynamics. The embedding is also used as the modeling basis for MPC optimization. For a freeway \zl{stretch} with unknown parameters and partial measurement on the freeway mainline, we provide sufficient conditions on the control horizon, cost functions, terminal sets of MPC, and inflow demand at the ramps such that the queue lengths in the closed-loop system remain bounded. The sufficient condition on the demand matches the necessary condition, thereby proving maximal throughput under the proposed controller. The result is strengthened to input-to-state stability when model parameters and demand are known. 
\zl{The stability analysis is conducted for the case of constant demand and unbounded on-ramps.} 
The \zl{closed-loop trajectory data} generated by the proposed controller is shown to facilitate finite time estimation of free-flow model parameters, i.e., free-flow speed and turning ratios.
Simulation results illustrate stability of the closed-loop system under the proposed controller with \zl{time-varying demand} and few  mainline measurements, for which the system becomes unstable under a well-known approach from the literature.
\zl{This indicates that the proposed controller renders higher throughput than the well-known approach, possibly using more computing resources.
}

\end{abstract}

\end{frontmatter}

\section{Introduction} 
Freeway traffic control is fundamental to the performance of transportation systems. The survey paper \cite{siri2021freeway} provides an overview of this problem. One of the main goals for freeway traffic control is to maintain as large throughput as possible. A commonly used traffic control method -- ramp metering -- regulates the entering flow to the mainline freeway from the on-ramps. An online optimization control approach called model predictive control (MPC) is one of the main paradigms in designing ramp metering controllers, e.g., see \cite{hegyi2005model,bellemans2006model}. 

Complete state information and traffic model parameters, such as parameters of the fundamental diagram and turning ratios, are often assumed to be available when formulating MPC ramp metering controllers, e.g., see \cite{hegyi2005model}. However, this is not always possible; thus, output-feedback and adaptive control are necessary. Output-feedback control deals with the case when the state is not directly measurable, and therefore, state estimation is typically pursued; adaptive control deals with the case when model parameters are unknown and need to be updated online. \zl{A natural approach to output-feedback or adaptive control, also known as certainty equivalence control, is to sequentially estimate states or parameters and then compute control actions. Although such a certainty equivalence approach is useful for linear systems, the closed-loop performance is not guaranteed for general nonlinear systems. When it comes to freeway traffic control, neither output feedback nor adaptive MPC has been studied rigorously with throughput guarantees, partly due to the nonlinearity of freeway traffic dynamics. As discussed in \cite{Mayne:14}, both output-feedback MPC and adaptive MPC have traditionally not been thoroughly studied for nonlinear systems, even separately. Only recently have there been some developments in this topic.}

For output-feedback MPC, \cite{copp2017simultaneous} develops a joint MPC-moving horizon estimation (MHE) method and provides sufficient conditions under which closed-loop system trajectories remain bounded; however, it is not clear how to verify the conditions for traffic dynamics. \cite{kohler2021robust} explores various options of estimation methods, including MHE and set-membership estimation; however, closed-loop system stability is not guaranteed. For adaptive MPC, a set-membership approach is proposed in \cite{adetola2009adaptive,adetola2011robust}, assuming that the dynamics are affine in parameters. A similar assumption is also made in some most recent studies, e.g., see \cite{lopez2019adaptive,sasfi2023robust}, while some of the most widely used traffic models, e.g., the Cell Transmission Model (CTM) \cite{daganzo1994cell,gomes2008behavior}, do not satisfy the assumption that the dynamics are affine in parameters. \zl{It can be seen that many of the output-feedback or adaptive MPC studies integrate set-membership estimation into controller design.}

\zl{Set-membership estimation computes set-valued estimates of states or parameters. The main challenge is to ensure that the estimated sets contain the corresponding true values. A closely related topic is reachability analysis, which computes the set of all reachable states starting from any initial condition, including the true initial condition. Therefore, the set of reachable states can serve as a set-valued state estimate, and tools from reachability analysis can be used to construct set-membership estimators, e.g., zonotopes \cite{rego2020guaranteed} and mixed-monotonicity \cite{coogan2020mixed,khajenejad2023tight}.
For traffic applications, set-membership estimation has been studied in \cite{kurzhanskiy2012guaranteed} to estimate an interval set of traffic states, i.e., upper and lower bounds on the density at every cell, with the guarantee that the actual state will lie in this set. Their estimator relies on the monotonicity of the underlying CTM, which does not hold under capacity drop phenomena \cite{kontorinaki2017first}}. In this study, 
We integrate the idea of set-membership estimation within control design and use the tool from mixed-monotonicity to overcome the assumptions on dynamics in previous studies. \zl{ Mixed-monotonicity holds for a general class of systems - bounded dynamics on any bounded set \cite{yang2019tight}. However, it has not been integrated into control design to the best of authors' knowledge.}

\zl{Our main contribution is the development of} a set-membership predictive control (Set-PC) approach to constructing output-feedback adaptive controller for ramp metering. This approach combines set-membership parameter and state estimation with MPC optimization. The set-membership estimation updates upper and lower bounds on unknown traffic states, traffic demand, and model parameters. The construction of upper- and lower-bound estimates relies on the mixed-monotone embedding of the traffic dynamics. The MPC optimization computes the control input based on embedding dynamics with the estimated bounds. 

    \begin{figure}
    \begin{subfigure}{0.2\textwidth}
        \centering
        \includegraphics[width=1.1\textwidth]{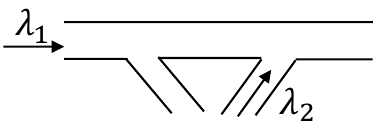}
        \caption{\sf Freeway stretch}
        \label{fig:freeway}
    \end{subfigure}
    \hspace{-55pt}
    \begin{subfigure}{0.5\textwidth}
        \centering
        \includegraphics[width=0.5\textwidth]{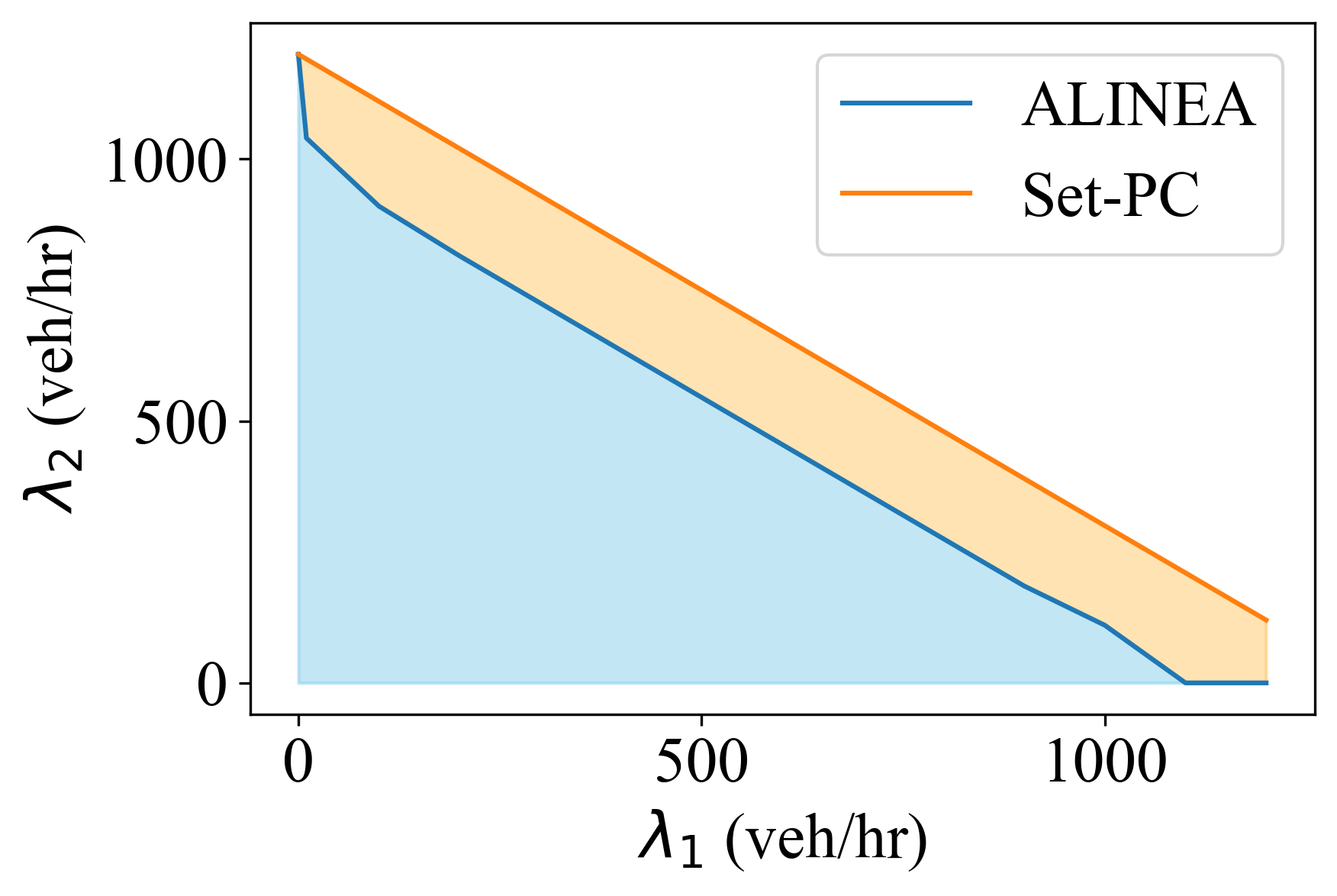}
        \caption{\sf Throughput region}
        \label{fig:fundamental-diagram}
    \end{subfigure}
    \caption{\sf \zl{For a freeway stretch in (a) with complete parameter and state information, the blue region in (b) represents the throughput region of ALINEA \cite{papageorgiou1991alinea} estimated from simulations. The region combining blue and orange is the throughput region of the proposed Set-PC controller. This combined region is maximal in the sense that no controller can stabilize the freeway under demands outside the region.}}
    \label{fig:maximal-throughput}
\end{figure}

The proposed Set-PC ramp metering controller renders maximal throughput to the freeway \zl{ with unbounded on-ramps. Throughput is measured in terms of the outflow from a freeway; it is equal to the inflow demand if queue lengths are bounded. The throughput \emph{region}, i.e., all the possible throughput provided by a given ramp metering controller, is characterized by the set of demands for which queue lengths are bounded under the given controller, starting from any initial condition. A controller is said to maximize throughput if the throughput region of the controller contains throughput region of every other controller. In this paper, we show that the proposed Set-PC controller maximizes throughput, see Fig.\ref{fig:maximal-throughput} for illustration. In other words, any demand for which there is a stabilizing\footnote{\zl{Here we refer to the weak notion of stability, i.e., boundedness of states.}} controller can be stabilized by the Set-PC controller.}

Specifically,
we provide sufficient conditions on the MPC time horizon, cost functions, terminal sets, and the demand, under which the on-ramp queue lengths remain bounded under the proposed controller, for the case when the unknown demand is constant. 
We also show that the \zl{closed-loops trajectory} generated by the proposed controller can be utilized to accurately estimate free-flow speed and turning ratios in finite time steps. 
The stability result is strengthened to input-to-state stability when the model parameters and demand are known. 
The sufficient condition on the demand for stability matches the necessary condition, thereby implying maximal throughput. 
It is important to note that we do not assume the system to be observable for our results. 
Indeed, our results suggest that maximal throughput can be achieved by utilizing \zl{measurements only from on-ramps and mainline entrance}. 
This is important because it has been noted that \zl{accurately estimating mainline state requires many mainline sensors} for freeway traffic dynamics, e.g., see \cite{munoz2006piecewise,nugroho2021should}.


The proposed Set-PC controller was first introduced in our conference paper \cite{li2023output}, where the output-feedback control was considered with known model parameters. This paper extends the formulation to the case when model parameters and traffic demand are unknown and provides new stability and parameter estimation results.

The rest of the paper is organized as follows. The problem setup is described in Section~\ref{sec:problem}. The Set-PC controller is described in Section~\ref{sec:mpc-set-membership}. Feasibility and sufficient conditions for stability are in Section~\ref{sec:stability-output-mpc}. 
Simulations are in Section~\ref{sec:simulations}, and concluding remarks are provided in Section~\ref{sec:conclusion}. Proofs of the technical results are collected in the Appendix.

We conclude this section by stating key notations to be used throughout the paper.  For integers $n_1$ and $n_2 \geq n_1$, we let \( [n_1:n_2]:= \{n_1, n_1+1, \ldots, n_2\} \). For brevity, $[1:n_1]$ will be denoted compactly as $[n_1]$. 
We shall use $x(t_1:t_2)$ to denote the sequence $\{x(t_1), x(t_1+1), \ldots, x(t_2)\}$.  $\mathbb{R}_{\geq0}$ and $\mathbb{R}_{>0}$ will denote the set of non-negative and positive reals, respectively. For vectors $x$ and $y$, we shall let $x \leq y$ imply componentwise inequalities. For a vector $x$, we shall denote its $i$-th component by $x_i$ or $[x]_i$. All matrices are denoted by uppercase letters in boldface to differ from numbers denoted by uppercase letters. $\mathbf{I}$ will denote the identity matrix whose dimension will be clear from the context. 

\zl{
\section{Literature Review}

\subsection{Model Predictive Control}
There is a vast literature on MPC in both control and transportation communities, see \cite{Mayne:14,siri2021freeway}. 
For nominal MPC, at each time step, control actions are computed by solving an optimization problem over a finite horizon starting from the known current state; the first of these control actions is implemented, and the process is repeated at the next time step, and so on. 
Stability issues for nominal MPC are covered in \cite{mayne2000constrained}. 
Input-to-state stability for MPC with disturbances is studied in \cite{lazar2008input}.
In output-feedback or adaptive setting, the nominal MPC is usually augmented with state or parameter estimators using sensor measurements, and state or parameter estimates are used in the optimization problem instead of the true values. 
Output-feedback MPC has been widely studied for linear time-invariant systems, e.g., see \cite{Mayne.Rakovic.ea:06}. 
However, very few studies have been conducted for nonlinear systems with performance guarantees, e.g., see \cite{copp2017simultaneous,kohler2021robust}.
Adaptive MPC has been studied for linear systems, e.g., see \cite{tanaskovic2014adaptive}, and for nonlinear systems with affine dynamics in parameters, e.g., see \cite{adetola2009adaptive,adetola2011robust,lopez2019adaptive,sasfi2023robust}.

\subsection{Ramp metering}
Ramp metering methods can be roughly classified into two categories: proportional-integral methods and optimization-based methods.
Proportional-integral methods, such as ALINEA \cite{papageorgiou1991alinea} and its variants \cite{wang2014local,frejo2018feed}, have been widely used in practice due to their simplicity and effectiveness \cite{papageorgiou2003review}. On the contrary, optimization-based methods, such as MPC, require more computational resources but can achieve better performance in terms of travel time \cite{hegyi2005model}.
Computational issues of MPC for ramp metering are investigated in \cite{muralidharan2015computationally}.

When traffic parameters are unknown or states are not directly measured, parameter or state estimation is usually augmented with ramp metering controllers. 
Typical sensors used in state estimation include loop detectors \cite{wang2005real} and connected vehicles \cite{bekiaris2016highway}.
An overview of freeway traffic state estimation is provided in \cite{seo2017traffic}.
For ALINEA, \cite{smaragdis2003series} proposed different modifications when different types of measurements are available;
\cite{smaragdis2004flow} estimated the key parameter, critical occupancy, in closed-loop with ALINEA. Different estimation methods have been used in closed loop with MPC, e.g., Luenberger estimation \cite{brandi2017model} and moving horizon estimation \cite{sirmatel2019nonlinear}.
To the best of authors' knowledge, set-membership estimation has not been used in closed loop with ramp metering controllers. 

}

\section{Problem formulation} 
\label{sec:problem}

\zl{\subsection{Traffic flow dynamics}}
\begin{figure}[htb!]
	\centering
	\includegraphics[scale=0.45]{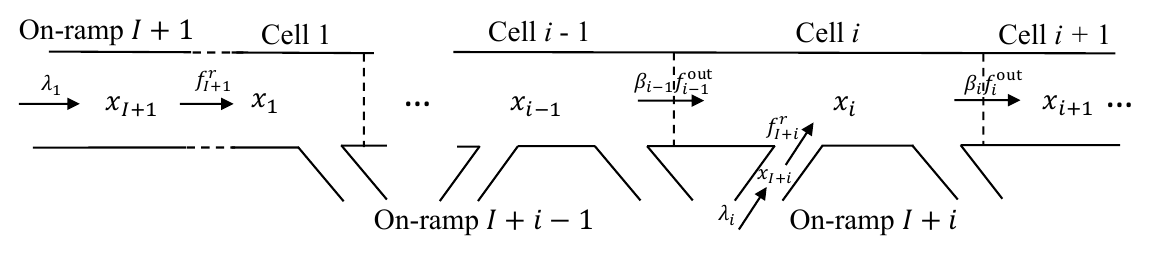}
	\caption{\sf A freeway stretch}
	\label{fig:network-sketch}
\end{figure} 
Consider a line freeway segment divided into $I$ \emph{mainline cells} indexed by $i=1,...,I$, each associated with one \emph{on-ramp} indexed by $i=I+1,\ldots,2I$, and one off-ramp; see Figure~\ref{fig:network-sketch} for illustration. 
\zl{The length of all mainline cells is normalized to one. Tables~\ref{tab:model-parameters-space}--\ref{tab:model-parameters-time} summarize the parameters and variables used in the dynamics.}
Let $x_i(t)\in\real_{\geq 0}$ denote the number of vehicles in mainline cell $i$ or on-ramp $I+i, i\in[I]$ at time $t$. Let $u_i(t)\in\real_{\geq 0}$ denote the desired metering rate at time $t$ of on-ramp $I+i$ connected to mainline cell $i$. $u_i(t)$ can be interpreted as the ramp-metering control \zl{policy} between ramp $I+i$ and mainline $i$. Let $\lambda_{i}(t)\in\real_{\geq 0}$ denote the traffic demand that enters from the on-ramp $I+i$ to the mainline $i$ at time $t$. Traffic demand can be seen as the exogenous input to the system.  Let $\fin_i\colon\real_{\geq 0}\times \real_{\geq 0}\times\real_{\geq 0}\to \real_{\geq 0}$ and $\fout_i\colon\real_{\geq 0}\times\real_{\geq 0}\times\real_{\geq 0}\to \real_{\geq 0}$ denote the inflow and outflow functions of cell $i$, respectively. 
Let $x(t):=\{x_i(t): \, i \in [2I]\}, u(t):=\{u_i(t): \, i \in [I]\}, \lambda(t):=\{\lambda_i(t): \, i \in [I]\}$ be the compact notations. Let $\theta\in\real_{\geq 0}^J$ denote the model parameter that will be specified later. 

We consider the traffic flow dynamics of the form
\begin{equation}\label{eq:cell-mass-balance}
	\begin{aligned}
			&x_i(t+1) \\ &\!=\! x_i(t) + \fin_i(x(t),u(t),\lambda(t);\theta) - \fout_i(x(t),u(t),\lambda(t);\theta) \\ & \!=:\! f_i(x(t),u(t),\lambda(t);\theta)  ,i\in[2I]
	\end{aligned}
\end{equation}
We use the Cell Transmission Model (CTM), see \cite{gomes2008behavior}, to specify the inflow function $\fin$ and outflow function $\fout$. According to the CTM, the mainline state $x_i$ is upper-bounded by jam density $\xjam_i$, i.e., $x_i\in[0,\xjam_i], i\in[I]$. We assume that on-ramps have infinite storage capacity, i.e., $x_i\in[0,+\infty),i\in[I+1,2I]$. The state space \zl{for both mainline and on-ramp states} is then defined as $\mc X(\theta):=[0, x^{\text{jam}}_1] \times \ldots \times [0, x^{\text{jam}}_I]\times [0,+\infty)^I$. The upper bound $\xjam$ for the state space is a model parameter in $\theta$. The dependence of $\mc X(\theta)$ on parameter $\theta$ is specifically on $\xjam$.
We assume that control $u_i$ is upper-bounded by a constant $\bar{u}_i$, i.e., $u_i\in[0, \bar{u}_i], i\in[I]$, and define $\mc U:=[0, \bar{u}_1] \times \ldots \times [0,\bar{u}_I]$. 

\begin{remark}\label{remark:ramp-outflow}
	The jam density $\xjam$ is the maximal number of vehicles allowed in each \zl{mainline} cell, which is determined by the length of each cell, the longitudinal distance between vehicles when they stop, \zl{and the number of lanes}.
	
	In practice, ramp metering control is implemented by setting the timings of the traffic signal at the end of an on-ramp. Therefore, the actual outflow from the on-ramp to the mainline at time $t$ is not always equal to control $u(t)$. This is because the outflow cannot exceed the existing queue length at on-ramps, and the resulting mainline state is bounded by jam density. Otherwise, the resulting state $x$ may not be in $\mc X(\theta)$.
\end{remark}

\begin{table}[t]
	\zl{
	\caption{Parameters and variables defined a point in time}
	\label{tab:model-parameters-space}
	\begin{tabular}{lll}
	\hline
	\textbf{Notation} & \textbf{Name} & \textbf{Unit} \\
	\hline
	$\xjam_i$ & Jam density & veh/cell \\
	$v_i$ & Free-flow speed & cell/time step \\
	$w_i$ & Congestion wave speed & cell/time step \\
	$x_i$ & Number of vehicles in cell $i$ & veh \\
	\hline
	\end{tabular}
	}
\end{table}
\begin{table}[t]
	\caption{Parameters and variables defined at a point in space}
	\label{tab:model-parameters-time}
	\begin{tabular}{lll}
	\hline
	\textbf{Notation} & \textbf{Name} & \textbf{Unit} \\
	\hline
	$\Cm_i$ & Capacity & veh/time step \\
	$\alpha_i$ & Capacity drop rate & dimensionless \\
	$\lambda_i$ & Demand & veh/time step \\
	$u_i$ & Metering rate & veh/time step \\
	$\beta_i$ & Turning ratio & dimensionless \\
	\hline
	\end{tabular}
\end{table}

Let $\beta_i \in (0,1), i\in[I-1]$ represent the fraction of outflow from cell $i$ that enters cell $i+1$; the rest of the $1-\beta_i$ fraction exits through the off-ramp; let $f_i^r\colon\real_{\geq 0}\times\real_{\geq 0}\times\real_{\geq 0}\to\real_{\geq 0}$ denote the actual outflow from on-ramp \zl{$I+i$} to mainline cell \zl{$i$}. \zl{Following Remark~\ref{remark:ramp-outflow}, the ramp outflow is constrained by ramp queue length and remaining space in mainline merge cell as follows:}
\begin{equation}\label{eq:ramp-outflow}
	\begin{aligned}
		&f_i^r(x,u,\lambda;\theta)  \\
		& := 
		\begin{cases}
			\min\{u_{i}, x_{I+i}+\lambda_{i},\xjam_{i} \\ \hspace{.5in} -(x_{i} - \fout_{i}(x,u,\lambda;\theta))\}, i=1  \\
			\min \{u_{i}, x_{I+i}+\lambda_{i}
			 \\ \hspace{0in} \xjam_{i}  - (x_{i}+  \beta_{i-1}\fout_{i-1}(x,u,\lambda;\theta) \\ \hspace{.8in}
			- \fout_{i}(x,u,\lambda;\theta))\}, i\in[2:I] 
		\end{cases}\\
	\end{aligned}
\end{equation}

Following the CTM, the inflow and outflow functions can be expressed as follows:
\begin{equation}\label{eq:define-inflow-outflow}
	\begin{aligned}
		&\fout_{i}(x,u,\lambda; \theta)\\ & \quad :=
		\begin{cases}
			\min\left\{d_i(x_i;\theta), s_{i+1}(x_{i+1};\theta)\right\}, & i \in [I-1] \\
			d_i(x_i;\theta), & i = I \\
		\end{cases}\\
		&\fout_{I+i}(x,u,\lambda; \theta) = f_i^r(x,u,\lambda;\theta), \qquad i\in[I] \\
		&\fin_i(x,u,\lambda;\theta)\\ & \quad :=
		\begin{cases}
			f_{i}^r(x,u,\lambda;\theta), & i = 1 \\
			\beta_{i-1} \fout_{i-1}(x,u;\theta) + f^r_{i}(x,u,\lambda;\theta), & i \in [2:I]		\end{cases} \\
		&\fin_{I+i}(x,u,\lambda;\theta) = \lambda_i, \qquad i\in[I]
	\end{aligned}
\end{equation}
where $d_i(x_i;\theta)$ and $s_i(x_i;\theta)$ are demand and supply functions of the mainline cell $i\in[I]$ that defines the fundamental diagram, see Figure~\ref{fig:fundamental-diagram}. 
\zl{For every mainline cell $i$, the demand function $d_i$ determines the maximum number of vehicles that can leave cell $i$ within one time step, and the supply function $s_i$ determines the maximum number of vehicles that can enter cell $i$ from the upstream mainline cell and on-ramps within one time step.}  
We consider the triangular fundamental diagram with capacity drop that uses piecewise linear demand and supply functions as follows:

\begin{equation}
	\label{eq:cell_demand_supply}
	\begin{aligned}
		d_i(x_i;\theta) &:= \min\left\{v_ix_i,\xi_i(x_i;\theta)\right\}, i\in[I] \\
		s_i(x_i;\theta) &:= \min\left\{\frac{w_{i}}{\beta_{i-1}}(\xjam_{i}-x_{i}), \Cm_{i-1}\right\}, i\in[2:I] \\
		\xi_i(x_i;\theta) &:= \begin{cases}
			\Cm_i, & x_i\leq \Cm_i/v_i\\
			\alpha_i\Cm_i, & x_i > \Cm_i/v_i
		\end{cases}, i\in[I]
	\end{aligned}
\end{equation}
where free-flow speed $v_i$, congestion wave speed $w_i$, jam density $\xjam_i$, capacity $\Cm_i$, and capacity drop rate $\alpha_i, i\in[I],$ are parameters in the traffic fundamental diagram shown in Figure~\ref{fig:fundamental-diagram}. Let $\xcrit_i:=\Cm_i/v_i$ denote the critical value of $x_i$ to indicate whether capacity drop occurs. 
We use the discontinuous function $\xi$ defined in \eqref{eq:cell_demand_supply} to model the capacity drop. Other \zl{formulations} of capacity drop can be found in \cite{kontorinaki2017first}.
\begin{figure}[htb!]
	\centering
	\includegraphics[scale = 0.7]{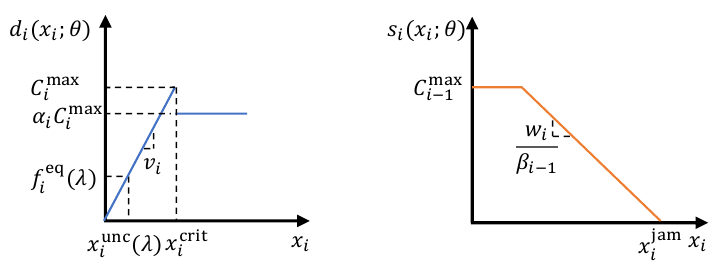}
	\caption{\sf Illustrations of the parameters of the fundamental diagram}
	\label{fig:fundamental-diagram}
\end{figure}
\zl{
\begin{remark}
	The CTM used in this paper is adapted from the Asymmetric Cell Transmission Model, see \cite{gomes2006optimal,gomes2008behavior}.
	The difference is that their model prioritizes on-ramps, meaning that on-ramp outflow can always be accommodated on the mainline. In contrast, the outflow model in \eqref{eq:ramp-outflow} prioritizes mainline pre-merging cells.
\end{remark}
}
We have completely defined the model \eqref{eq:cell-mass-balance} where $\theta$ includes all the parameters in the fundamental diagram and turning ratios, i.e.,
	$$\theta^T:=\begin{bmatrix}
		\beta^T & v^T & w^T & \xjamT  & \CmT & \alpha^T
	\end{bmatrix}$$ We denote the parameter space as $\theta\in\Theta\subset\real_{\geq 0}^J$ where $J$ is the total number of parameters.
\begin{remark}\label{remark:CFL} The CTM is a numerical scheme for solving the Lighthill-Whitham-Richards model \cite{lighthill1955kinematic,richards1956shock}, which is a partial differential equation for traffic flow dynamics. 
	For the stability of the numerical method, the Courant–Friedrichs–Lewy (CFL) condition\cite{courant1928} implies that\/ $w_i, v_i\leq 1, i\in[I]$\@ for \eqref{eq:cell-mass-balance}. Under these conditions, the dynamics \eqref{eq:cell-mass-balance} is well-defined in $\mc X(\theta)$.
\end{remark}



\zl{\subsection{Equilibrium analysis}}
Let
$\mathbf{R}$ be an $I \times I$ matrix and let \zl{$\mathbf{R}_{i,j}$ denote the entry at row $i$ and column $j$}. We define the routing matrix $\mathbf{R}$ such that $\mathbf{R}_{i,i-1}=\beta_{i-1}$ for all $i \in [2:I]$, and all other entries being zero. 
%
 Let $x^m$ and $x^r$ denote the mainline and on-ramp states, i.e., $x^m := \{x_i:i\in[I]\}, x^r:=\{x_{I+i}:i\in[I]\}$, respectively. Let $\foutm$ denote the outflow of the mainline cells, i.e., $\foutm:=\{\fout_i:i\in[I]\}$. To analyze the equilibrium of the dynamics  \eqref{eq:cell-mass-balance}, we consider the case when \zl{$\fout_{I+i}(x, u,\lambda;\theta)= u_{i},i\in[I]$, }which implies that the actual ramp outflow is equal to the control $u$ that ensures $x\in\mc X(\theta)$. Then, \eqref{eq:cell-mass-balance} can be compactly written as:
\begin{equation}
	\begin{aligned}\label{eq:cell_compact_dynamic}
		x^m(t+1) &= x^m(t) - (\IR)\foutm(x(t);\theta) + u(t) \\
		x^r(t+1) &= x^{r}(t) - u(t) + \lambda(t)
	\end{aligned}
\end{equation}

In the uncontrolled case with constant demand, i.e., when $\lambda(t)\equiv \lambda$ and $u(t) \equiv \lambda$, \zl{from \eqref{eq:cell_compact_dynamic} mainline equilibrium flow is determined by \footnote{$(\IR)^{-1}=\mathbf{I} + \mathbf{R} + \mathbf{R}^2 +\ldots$ since all the entries in $\mathbf{R}$ are non-negative and strictly less than one.
} 
\begin{equation}\label{eq:equilibrium-flow}
	\foutm(x)=(\IR)^{-1}\lambda=:\feq(\lambda)
\end{equation}
Since all entries in $(\IR)^{-1}$ are nonnegative \footnote{The same reason above},  $\feq(\lambda)$ is monotonically increasing with respect to $\lambda$.}
It is natural to assume that $\lambda$ is sufficiently small so that 
\begin{equation}
	\label{eq:max-flow}
	\feq_i(\lambda) \leq \Cm_i, \quad i \in [I]
\end{equation}
i.e., the equilibrium flow $\feq(\lambda)$ in the uncontrolled case does not exceed the flow capacity of the fundamental diagram on mainline cells. 
We say that the mainline is uncongested at time $t$ if $x_i^m(t)\leq \xcrit_i$ for all $i\in[I]$. 
Without loss of generality, we assume that $\xcrit_i = \Cm_i/v_i=\max\{x_i:\foutm_i(x;\theta) = v_ix_i\}, i\in[I]$, which implies $\foutm_i(x,\theta) = v_ix_i$ if $x_i\leq \xcrit_i$. If $\Cm_i/v_i > \max\{x_i:\foutm_i(x;\theta) = v_ix_i\}$, we can always decrease the value of $\Cm_i$ to make the two sides equal without affecting the dynamics. 
Then, for every $\lambda$ satisfying \eqref{eq:max-flow}, there exists a unique \emph{uncongested} equilibrium state $\xunc(\lambda)$ on the mainline since the equilibrium flow is unique and $\feq_i(\lambda) = v_i \xunc_i(\lambda)$ for all $i \in [I]$ (see Figure~\ref{fig:fundamental-diagram} for illustration).


%

\zl{\subsection{Output-feedback adaptive control}}
We consider the following output model:
\begin{equation}\label{eq:output_measurements}
	\underbrace{\begin{bmatrix}y^m(t) \\ y^r(t) \end{bmatrix}}_{y(t)} = \begin{bmatrix}\mathbf{C} & \mathbf{0} \\
		\mathbf{0} & \mathbf{I} \end{bmatrix} \begin{bmatrix} x^m(t) \\ x^r(t) \end{bmatrix} 
\end{equation}
for appropriate, not necessarily invertible matrix $\mathbf{C}$. That is, we assume that we have perfect information about the states of the ramps \zl{and mainline entrance, but not necessarily for the remaining mainline sections.} 

\begin{example}
	\label{ex:C-matrix}
	\zl{
	\emph{Space occupancy} is a commonly available measurement, e.g., through paired loop detectors.}
	\cite{cassidy1997relation} shows that space occupancy in each mainline cell $i$ can be calculated as $x^m_i\times\text{average vehicle length }/\text{ length of cell $i$}$. Therefore, when occupancy measurements are available,\/ $\mathbf{C}$ in \eqref{eq:output_measurements} is a diagonal matrix, with zero diagonal entries corresponding to missing detectors on the corresponding cells.
	%
\end{example}
Our objective in this paper is to design an output-feedback adaptive controller which renders maximum throughput to the freeway stretch. 
We make the following assumption on the demand, which turns out to be necessary for the constant demand case:
\begin{assumption}\label{ass:demand-average}
	The sequence of time-varying demand $\{\lambda(t)\}_{t=0}^{\infty}$ satisfies 
    \begin{equation}\label{eq:demand-necessary}
        \zl{\limsup_{\tau\to\infty}\ (\IR)^{-1}\frac{1}{\tau}\sum_{t=0}^\tau \lambda(t)\leq\Cm}
    \end{equation}
\end{assumption}
\begin{remark}
	Assumption~\ref{ass:demand-average} requires that the time average of demand satisfies \eqref{eq:max-flow}. In the case of constant demand, i.e., $\lambda(t)\equiv\lambda$, \eqref{eq:demand-necessary} reduces to \eqref{eq:max-flow}, and this is a necessary condition for any controller to achieve bounded queue lengths. \cite{gomes2008behavior} shows that if \eqref{eq:max-flow} is not satisfied, then no equilibrium exists, and the queue lengths grow unbounded.
\end{remark}
\zl{
\begin{remark}
	In the CTM variant used in this paper~ \cite{gomes2008behavior}, the flow entering the mainline is assumed to be controlled.
	This is not necessarily the case in practice.
	When the flow entering the mainline is not controlled, then 
	the necessary condition \eqref{eq:demand-necessary} is not tight, and a subset of demand satisfying \eqref{eq:demand-necessary} needs to be considered.
	In the case of constant demand, this subset is constructed by adding an additional constraint that the mainline entering flow $\lambda_0$ does not exceed $\min_{i\in{I}}\alpha_i\Cm_i$.
	The subset of $\lambda$ is necessary for any controller to achieve bounded queue lengths when the mainline entering flow is not controlled.
	A simple counterexample is that if $\lambda_0>\min_{i\in{I}}\alpha_i\Cm_i$ and initial conditions are congested, 
	then no control can steer the system to the uncongested region. Then, there always exists a mainline cell for which the outflow exceeds the dropped capacity under any controllers, and thus total queue length grows unbounded.
	The analysis throughout the paper applies to this subset.
\end{remark}}

The output-feedback adaptive control that we shall design will ensure that if the constant demand satisfies the necessary condition in \eqref{eq:demand-necessary}, then the queue length remains bounded, starting from any initial condition $x(0)\in\mc X(\theta)$.
\zl{The necessary condition \eqref{eq:demand-necessary} provides an upper bound on demand $\lambda$. The set of constant demand satisfying \eqref{eq:demand-necessary} contains  throughput region of any controller}. \zl{Our controller will be shown to maximize throughput since the controller stabilizes the system under all constant demand satisfying \eqref{eq:demand-necessary}.}

Considering that the  \emph{model predictive control} (MPC) approach is widely used in ramp metering, e.g., see \cite{hegyi2005model}, we shall extend it to output feedback adaptive control and establish its maximal throughput property. For the MPC approach, it is typically of interest to design a control signal $\{u(t)\colon \, t\geq 0\}$ to  
minimize a cost $\sum_{t=0}^{\infty} \zl{\ell}(x(t),u(t))$, subject to dynamics \eqref{eq:cell-mass-balance}, $x(t)\in\mc X(\theta), u(t)\in\mc U$ for a given non-negative function $\zl{\ell}(\cdot,\cdot)$. Instead of solving this infinite horizon problem, the MPC approach recursively solves a related finite horizon problem, which at time $t$ is:
\begin{equation}
	\label{eq:finite-horizon-cost}
	\sum_{k=t}^{t+T-1} \zl{\ell}(x(k), u(k)) + V_f(x(t+T))
\end{equation}
subject to dynamics \eqref{eq:cell-mass-balance}, $x(k)\in\mc X(\theta), u(k)\in\mc U,$  and \emph{terminal constraint} $x(t+T)\in \mc X_f$, where $T$ is the \emph{forward} horizon, and $V_f(\cdot)$ is the terminal cost that is non-negative. Let $\{\hat{u}^*(t), \ldots, \hat{u}^*(t+T-1)\}$ denote an optimal solution to \eqref{eq:finite-horizon-cost}, we set $u(t)$ to be equal to $\hat{u}^*(t)$, then \eqref{eq:finite-horizon-cost} is re-solved at $t+1$ to similarly obtain $u(t+1)$, and so on. 
\begin{remark}\label{remark:cost-function}
	\zl{
	A standard choice of the running cost is \emph{total time spent}: $\ell(x(t),u(t)) = \sum_{i\in[I]}x_i(t) + x_{I+i}(t)$ that represents total number of vehicles at time $t$. Summing $\ell$ over time can be interpreted as the total time spent on the network by all vehicles. Any linear combination of the costs can be used to trade-off between the total time spent on different mainline cells and ramps.
	}
\end{remark}

In standard state-feedback MPC, $x(t),\lambda(t)$, and $\theta$ are assumed to be known when solving \eqref{eq:finite-horizon-cost}. In this paper, we are rather interested in the setting where $x(t),\lambda(t)$, and $\theta$ are unknown, and the controller is only given the past input-output data $u(t-1),\ldots,u(t-L)$ and $y(t), y(t-1), \ldots, y(t-L)$ with \emph{backward} horizon $L$ when solving \eqref{eq:finite-horizon-cost} at each time $t$. A natural approach is to augment MPC with a simultaneous \emph{state} and \emph{parameter} estimation component. We approach the problem from the perspective of \emph{set-membership estimation}, e.g., see \cite{kurzhanskiy2012guaranteed} in the context of traffic dynamics, and accordingly pursue the output-feedback adaptive MPC approach. We name the approach as set-membership predictive control (Set-PC). The following section elaborates on the Set-PC controller. 

\section{The Set-PC controller}
\zl{\subsection{General framework}}
\label{sec:mpc-set-membership}
\zl{We aim to estimate interval sets that contain true values of states, exogenous demand, and parameters. To achieve this, we embed the dynamics $f$ into a twice-dimensional space, whose state variable is denoted as $\xlift^T = \begin{bmatrix}
\bar{x}^T & \ubar{x}^T
\end{bmatrix}$. We further augment the state with demand $\lambdalift^T = \begin{bmatrix}
	\lambdaup^T & \lambdalb^T
\end{bmatrix}$ and parameters $\thetalift^T = \begin{bmatrix}
	\thetaup^T & \thetalb^T
\end{bmatrix}$ in the lifted space. The lifted state space is denoted as $\Xlift(\thetalift) := \mc X(\bar{\theta}) \times \mc X(\ubar{\theta})$.
The dynamics in this lifted space is given by
\begin{equation}\label{eq:lift-dynamics}
	\begin{aligned}
		\xlift(t+1) &= \Flift(\xlift(t),u(t),\lambdalift(t);\thetalift(t)) \\
	\end{aligned}
\end{equation}
where $\Flift$ is the embedding dynamics for the state.
The dynamics is to be chosen such that the lifted variables are the extremes of interval uncertainty of the corresponding quantities in the original (lower-dimensional) space.
For example, $[\ubar{x}_i,\bar{x}_i]$ is the interval uncertainty for $x_i$. 

Similarly, for demand and parameters, we design $\Ulift^{\lambda}$ and $\Ulift^{\theta}$ as the estimator dynamics to characterize the corresponding interval uncertainties. 
Also, we wish to incorporate the past $L$ time steps of input-output data for parameter estimates. Let $\xlift_L(t)$ and $y_L(t)$ denote the sequences of lifted states and outputs from time $t-L$ to $t$, respectively; let $u_L(t-1)$ and $\lambdalift_L(t-1)$ denote the corresponding sequences from $t-L$ to $t-1$.
The estimator dynamics can be written as:
\begin{equation}
	\begin{aligned}
		\lambdalift(t) &= \Ulift^{\lambda}(\xlift_L(t),u_L(t-1),y_L(t),\lambdalift_L(t-1);\thetalift(t-1)) \\
		\thetalift(t) &= \Ulift^{\theta}(\xlift_L(t),u_L(t-1),y_L(t),\lambdalift_L(t-1);\thetalift(t-1))
	\end{aligned}
\end{equation} 
In other words, we aim to design $\Flift,\Ulift^{\lambda}$, and $\Ulift^\theta$
such that, if the initial condition $(\xlift(0),\thetalift(0), \lambdalift(0))$ characterizes the interval uncertainty of the true initial values $(x(0),\theta,\lambda(0))$, then the lifted state $(\xlift(t),\thetalift(t),\lambdalift(t))$ characterizes the interval uncertainty of the true values $(x(t),\theta,\lambda(t))$ for all $t\geq0$.

Let $\kappa(\xlift,\lambdalift,\thetalift)$ denote an MPC control law that maps the lifted state, demand, and parameters to control input $u$.
We consider the following Set-PC controller, see Fig.~\ref{fig:control-flow-chart}, which combines $\kappa$ with state, demand, and parameter estimates, starting with some initial estimates $\tilde{\xlift}(0),\lambdalift(-1)$, and $\thetalift(-1)$:
\begin{figure}[hbt]
	\centering
	\includegraphics[width=0.43\textwidth]{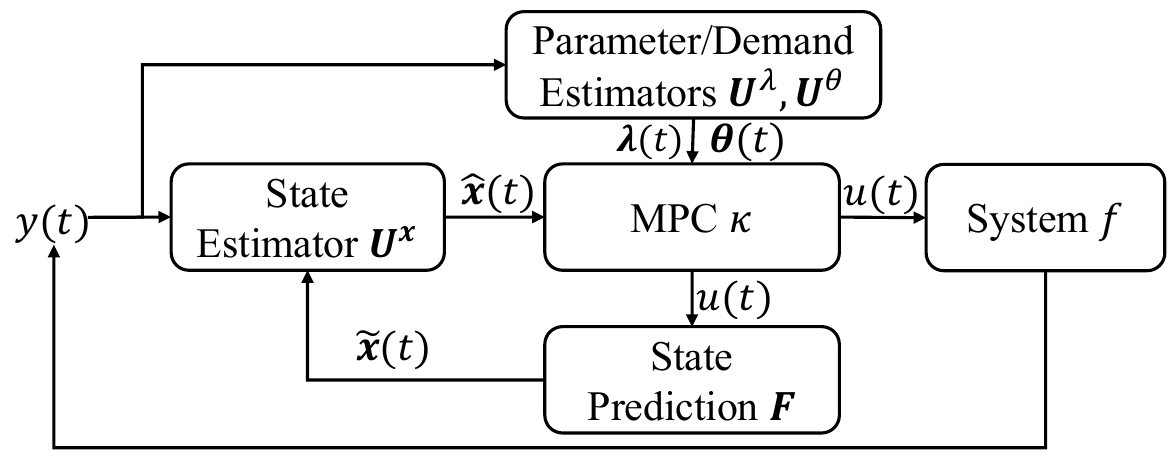}
	\caption{\sf Set-PC Controller}
	\label{fig:control-flow-chart}
\end{figure}
\begin{subequations}
	\label{eq:total-framework}
	\begin{align}
	&\hat{\xlift}(t)  = \Ulift^x(\tilde{\xlift}(t), y(t))  \nonumber \\ & \hspace{1in} \text{(State Estimate Update)} \label{eq:state-update}\\
	&\thetalift(t) = \Ulift^{\theta}(\hat{\xlift}_L(t),u_L(t-1),y_L(t),\lambdalift_L(t-1);\thetalift(t-1))
	\nonumber \\ & \hspace{1in} \text{(Parameter Estimate Update)} \label{eq:parameter-update}\\
	&\lambdalift(t) = \Ulift^{\lambda}(\hat{\xlift}_L(t),u_L(t-1),y_L(t),\lambdalift_L(t-1);\thetalift(t-1)) \nonumber \\ & \hspace{1in} \text{(Demand Estimate Update)} \label{eq:disturbance-update} \\
	&u(t) = \kappa(\hat{\xlift}(t),\lambdalift(t),\thetalift(t)) \nonumber \\ & \hspace{1in} \text{(MPC)} \label{eq:control-computation}\\
	&\tilde{\xlift}(t+1) = \Flift(\hat{\xlift}(t), u(t), \lambdalift(t); \thetalift(t)) \nonumber \\ & 
	\hspace{1in} \text{(State Prediction)} \label{eq:predict}
	\end{align}
\end{subequations}
\begin{remark}
	At initial time $t=0$, we implicitly assume that the past $L$ time steps of data are available for parameter estimation, i.e., $\xlift_L(0),u_L(-1),y_L(0),\lambdalift_L(-1)$ are given. This can be obtained by setting arbitrary input $u$ for $L$ time steps before implementing \eqref{eq:total-framework} at\/ $t=0$. Throughout the paper, theoretical results apply to any $L\geq1$ and any input sequence before $t=0$. 
\end{remark}
The \emph{predict-update} structure of \eqref{eq:total-framework} is reminiscent of popular filtering algorithms, in our case interspersed with control and parameter estimation. The following assumption is made such that initial estimates characterize the interval uncertainty of the true initial values:
\begin{assumption}\label{assu:initial-condition}
	The given initial estimates $\tilde{\xlift}(0),\thetalift(-1)$, and $\lambdalift(-1)$ satisfy that $\tilde{\ubar{x}}(0)\leq x(0)\leq\tilde{\bar{x}}(0), \thetalb(-1)\leq\theta\leq\thetaup(-1)$, and\/ $\lambdalb(-1)\leq\lambda(t)\leq\lambdaup(-1)$\@ for all $t\geq0$. Additionally, $\bar{x}^{\text{jam}}(-1) = \ubar{x}^{\text{jam}}(-1)$.
		
\end{assumption}
\begin{remark}
	Assumption~\ref{assu:initial-condition} requires initial estimates to be upper and lower bounds of unknown quantities, and jam density is known, which is consistent with the requirements in \cite{kurzhanskiy2012guaranteed}. Such estimates can be obtained from historical data such as PeMS \cite{chen2001freeway}, and jam density can be reliably estimated as opposed to other parameters \cite{kurzhanskiy2012guaranteed}. 
\end{remark}
In the remaining part of this section, we will show how to design $\Flift$ in Section~\ref{sec:mixed-monotone}, $\Ulift^x,\Ulift^{\theta}$, and $\Ulift^{\lambda}$ in Section~\ref{sec:state-demand-parameter-estimates}, and $\kappa$ in Section~\ref{sec:mpc-control-law} such that $(\xlift(t),\thetalift(t),\lambdalift(t))$ characterizes the interval uncertainty for all $t\geq0$.
}

\zl{
\subsection{Mixed-monotone embedding}
\label{sec:mixed-monotone}
The embedding dynamics $\Flift$ needs to satisfy that $\xlift(t+1)$ updated from \eqref{eq:lift-dynamics} characterizes the interval uncertainty of $x(t+1)$ if $\xlift(t)$ characterizes the interval uncertainty of $x(t)$.
The literature on \emph{mixed-monotonicity}, e.g., see \cite{enciso2006nonmonotone}, suggests $\Flift(\xlift,u,\lambdalift;\thetalift)=\begin{bmatrix} F(\xlift,u,\lambdalift;\thetalift) \\  F(\xlift^{\dagger},u,\lambdalift^{\dagger};\thetalift^{\dagger})\end{bmatrix}$,\\ with $\xlift^{\dagger}=\begin{bmatrix}\bar{x} \\ \ubar{x}\end{bmatrix}^{\dagger}:=\begin{bmatrix}\ubar{x} \\ \bar{x} \end{bmatrix}$, and $F$ having the following monotonicity properties:
\begin{enumerate}
\vspace{-0.1in}
\item[(M1)] $F\left(\begin{bmatrix} x \\ x \end{bmatrix},u,\begin{bmatrix} \lambda \\ \lambda \\ \end{bmatrix} ;\begin{bmatrix} \theta \\ \theta \end{bmatrix} \right)=f(x,u,\lambda;\theta)$
\item[(M2)] $F\left(\begin{bmatrix} x \\ z \end{bmatrix},u,\begin{bmatrix} \lambda \\ \phi \end{bmatrix};\begin{bmatrix} \theta \\ \eta \end{bmatrix}\right) $ \ $\leq F\left(\begin{bmatrix} \tilde{x} \\ z \end{bmatrix},u,\begin{bmatrix} \tilde{\lambda} \\ \phi \end{bmatrix};\begin{bmatrix} \tilde{\theta} \\ \eta \end{bmatrix}\right)$ if $x \leq \tilde{x}$, $\lambda \leq \tilde{\lambda}$ and $\theta \leq \tilde{\theta}$
\item[(M3)] $F\left(\begin{bmatrix} x \\ z \end{bmatrix},u,\begin{bmatrix} \lambda \\ \phi \end{bmatrix};\begin{bmatrix} \theta \\ \eta \end{bmatrix}\right) \leq F\left(\begin{bmatrix} x \\ \tilde{z} \end{bmatrix},u,\begin{bmatrix} \lambda \\ \tilde{\phi} \end{bmatrix};\begin{bmatrix} \theta \\ \tilde{\eta} \end{bmatrix}\right)$ if $z \geq \tilde{z}$, $\phi \geq \tilde{\phi}$ and $\eta \geq \tilde{\eta}$
\end{enumerate} 
\begin{remark}
	The three properties are such that $F$ decomposes the dynamics into nonincreasing and nondecreasing components.
	Such \emph{mixed-monotonicity} mappings have attracted a lot of interest recently in the context of reachability analysis, e.g., see \cite{coogan2020mixed}.
\end{remark} 
A mapping $F$ with the above monotonicity properties (M1)-(M3) is known to exist for any $f$ that is bounded on any bounded sets \cite{yang2019tight,coogan2020mixed}. However, it is hard to find analytical $F$ following the expressions in \cite{yang2019tight,coogan2020mixed} that formulate $F$ as optimization problems. 
Using domain knowledge in our problem, we propose the following mixed-monotone mapping:
\begin{equation}\label{eq:mixed-monotone-decomposition-function}
	F(\xlift,u,\lambdalift;\thetalift)=\bar{x}+\Fin(\xlift,u,\lambdalift;\thetalift)-\Fout(\xlift,u,\lambda;\thetalift)
\end{equation}
with $\Fin_i(\xlift,u,\lambdalift;\thetalift)=u_i$ if $i=1$ and $=u_i+ \bar{\beta}_{i-1} \allowbreak \min\{\tilde{d}_{i-1}(\bar{x}_{i-1},\ubar{x}_{i-1};\thetaup,\thetalb),\tilde{s}_i(\bar{x}_i;\thetaup,\thetaup)\}$ if $i\in[2:I]$; 
$\Fin_{I+i}(\xlift,u,\lambdalift;\thetalift)=\lambdaup_i$ for all $i\in[I]$; 
$\Fout_i(\xlift,u,\lambdalift;\thetalift)=\min\{\tilde{d}_i(\bar{x}_i,\bar{x}_i;\thetalb,\thetaup), \tilde{s}_{i+1}(\bar{x}_{i+1};\thetalb,\thetaup)\}$ if $i\in[I-1]$ and $=d_i(\bar{x}_i;\thetalb)$ if $i=I$;
and $\Fout_{I+i}(\xlift,u,\lambdalift;\thetalift)=u_i$ for all $i\in[I]$, where $\tilde{d}_i(x_i, z_i;\theta,\tilde{\theta}) := \min\{v_ix_i, \xi(z_i;\theta,\tilde{\theta})\}, i\in[I]$ with $\xi(z_i;\theta,\tilde{\theta}) =\Cm_i$ if $z_i\leq \Cm_i/\tilde{v}_i$ and $=\alpha_i \Cm_i$ otherwise, and $\tilde{s}_i(x_i;\theta,\tilde{\theta}) := \max\{0,\min\{w_i/\tilde{\beta}_{i-1}(\xjam_i - x_i),\Cm_{i-1}\}\}$ for all $i\in[2:I]$.
\begin{proposition}\label{prop:mixed-monotone}
The mixed-monotone mapping \eqref{eq:mixed-monotone-decomposition-function} satisfies properties $(\emph{M1})-(\emph{M3})$ for dynamics $f$ in \eqref{eq:cell-mass-balance} if $\fout_{I+i}(x,u,\lambda;\theta)=u_{i}, i\in[I]$.
\end{proposition}
\begin{remark}
	The dynamics $f$ in \eqref{eq:cell-mass-balance} is equivalent to \eqref{eq:cell_compact_dynamic} if $\fout_{I+i}(x,u,\lambda;\theta)=u_{i}, i\in[I]$. 
	The MPC optimization will be designed to ensure that computed control\/ $u$ satisfies this condition.
\end{remark}
}

\zl{
\subsection{State, demand, and parameter estimation}
\label{sec:state-demand-parameter-estimates}
The estimates
$\hat{\xlift}(t)$, $\lambdalift(t)$, and $\thetalift(t)$ in \eqref{eq:total-framework} need to characterize the interval uncertainty of $x(t)$, $\lambda(t)$, and $\theta$ if $\tilde{\xlift}(t)$, $\lambdalift(t-1)$, and $\thetalift(t-1)$ characterize the interval uncertainty of $x(t)$, $\lambda(t)$, and $\theta$. For state estimator $\Ulift^x$, we make the following assumption:
}
\begin{assumption}\label{ass:C-matrix}
	\zl{Matrix\/ $\mathbf{C}$ in the output model \eqref{eq:output_measurements} is diagonal.}
\end{assumption}
\begin{remark}
	Example~\ref{ex:C-matrix} shows that Assumption~\ref{ass:C-matrix} is satisfied when space occupancy measurements are available. However, there are no minimum requirements on the availability of mainline measurements, i.e., on the number of nonzero elements in $\mathbf{C}$.
\end{remark} 

\zl{
For diagonal $\mathbf{C}$, a natural state estimator is $\Ulift^x(\xlift,y)=\begin{bmatrix}
	U^x(\bar{x},y) \\ U^x(\ubar{x},y)
\end{bmatrix}$, where for all $i\in[I], U^x_i(x,y) = y_i / \mathbf{C}_{i,i}$ if $\mathbf{C}_{i,i}\neq0$ and $=x_i$ otherwise; $U^x_{I+i}(x,y) = y_{I+i},i\in[I]$. 

The parameter estimator $\Ulift^\theta$ is defined in terms of appropriate optimization problems. 
Specifically, let $\Ulift^\theta = \begin{bmatrix}
	U^{\theta,\max}(\xlift_L(t),u_L(t-1),y_L(t),\lambdalift_L(t-1);\thetalift(t-1)) \\ U^{\theta,\min}(\xlift_L(t),u_L(t-1),y_L(t),\lambdalift_L(t-1);\thetalift(t-1))
\end{bmatrix}$, where for all $j\in[J]$,
\begin{equation}\label{eq:U-theta-max}
	\begin{aligned}
		&U^{\theta,\max}_j(\xlift_L(t),u_L(t-1),y_L(t),\lambdalift_L(t-1);\thetalift(t-1)) \\
		= & \max_{\thetalb(t-1)\leq\theta\leq\thetaup(t-1)} \theta_j\quad \text{s.t. } \eqref{eq:cell-mass-balance},\eqref{eq:output_measurements} \text{ from $t-L$ to $t$ with} \\ & \ubar{x}(t-L)\leq x(t-L)\leq\bar{x}(t-L)\\
	\end{aligned}
\end{equation}
and $U^{\theta,\min}$ is defined similarly with $\max$ replaced by $\min$ in \eqref{eq:U-theta-max}.
}
\begin{remark}\label{remark:mixed-integer-bilinear}
	The optimization problems in \eqref{eq:U-theta-max} \zl{can be reformulated as} mixed-integer \zl{bilinear programs (MIBLP), e.g., see \cite{fischetti2020branch}. MIBLP are optimization problems with integer variables and bilinear costs or constraints.
	In (12), bilinear constraints appear in multiplications of parameters in dynamics \eqref{eq:cell-mass-balance} and integer variables result from reformulating capacity drop and pointwise minimum in \eqref{eq:cell_demand_supply} as constraints with integers, see \cite{muralidharan2015computationally}.}
	Existing solvers such as \texttt{Gurobi} \cite{gurobi} can solve this class of problems efficiently. 
\end{remark}
\zl{
Unlike constant parameter $\theta$, demand $\lambda(t)$ is time-varying, and the optimization-based estimator in \eqref{eq:U-theta-max} cannot be used to characterize uncertainty interval of $\lambda(t)$. Instead, we use a simple rule\/ $ \Ulift^\lambda(\xlift,u,\lambdalift;\thetalift) = \lambdalift$.
}
\begin{remark}
	 \zl{The estimator\/ $\Ulift^\lambda$ keeps $\lambdaup(t)$ and $\lambdalb(t)$ the same over time, i.e., $\lambdaup(t)=\lambdaup(0)$ and $\lambdalb(t)=\lambdalb(0)$ for all $t\geq0$}. This is sufficient for the stability of the closed-loop system in Theorem~\ref{theo:stability-boundedness} under constant demand and in Section~\ref{sec:simulation-unknown} for periodic demand. Other predictors can be considered to handle complex demand profiles. This would be pursued in future work.
\end{remark}
\zl{
\subsection{MPC control law}
\label{sec:mpc-control-law}
The MPC problem in the lifted space is that,  given estimates $\hat{\xlift}(t),\lambdalift(t)$, and $\thetalift(t)$, solve:

\begin{equation}\label{eq:mpc-opt-adaptive}
	\begin{aligned}
		&\min_{\hat{u}(0|t), \ldots, \hat{u}(T-1|t)} \quad \sum_{k=0}^{T-1}\ell(\hat{\bar{x}}(k|t), \hat{u}(k|t)) + V_f(\hat{\bar{x}}(T|t))\\
		&\text{s.t. } \\ 
		&\quad \hat{\xlift}(k+1|t) = \Flift(\hat{\xlift}(k|t),\hat{u}(k|t),\lambdalift(t);\thetalift(t))
		\\ & \hspace{2.35in} k\in[0\colon T-1] \\
		&  \quad \hat{\xlift}(0|t) = \hat{\xlift}(t),  \quad \hat{\xlift}(T|t) \in \Xflift
		\\
		& \quad \hat{\xlift}(k|t)\in\Xlift(\thetalift(t)),\hat{u}(k|t) \in \mc U, \hspace{0.5in} k\in[0\colon T-1] \\
	\end{aligned}
\end{equation}
where terminal set $\Xflift$ is set be box constraints, i.e., $\Xflift = \mc X_f\times \mc X_f$ with $\mc X_f:= \{x:0\leq x \leq \xter\}$ and a user-defined parameter $\xter\geq 0$. Given $\hat{\xlift}(t), \lambdalift(t)$ and $\thetalift(t)$, we denote the optimization \eqref{eq:mpc-opt-adaptive} as $\mathbbm{P}(\hat{\xlift}(t),\lambdalift(t),\thetalift(t))$.
Let $\{\hat{u}^*(0|t),\ldots,\hat{u}^*(T-1|t)\}$ be an optimal solution to $\mathbbm{P}(\hat{\xlift}(t),\lambdalift(t),\thetalift(t))$, we set $\kappa(\hat{\xlift}(t), \lambdalift(t),\thetalift(t))=\hat{u}^*(0|t)$.
\begin{remark}\label{remark:mixed-integer-linear}
	The optimization in \eqref{eq:mpc-opt-adaptive} is a mixed-integer linear program following the same reasoning as in Remark~\ref{remark:mixed-integer-bilinear}. Unlike the optimization in \eqref{eq:U-theta-max}, bilinearity does not appear in \eqref{eq:mpc-opt-adaptive} since parameter $\thetalift(t)$ is given.
\end{remark}
}

\section{Properties of the Set-PC controller}
\label{sec:stability-output-mpc}

\zl{
We have specified the design of each component in the Set-PC controller in \eqref{eq:total-framework}. 
We now state theoretical guarantees of the Set-PC controller, including guarantees on set-valued estimation, recursive feasibility, stability, and convergence of parameter estimation.

\subsection{Guaranteed set-membership estimation}
The following result shows that the Set-PC controller generates estimates that characterize the interval uncertainty of true states, parameters, and demand.
\begin{lemma}\label{lemm:set-membership}
	Let Assumptions~\ref{assu:initial-condition}--\ref{ass:C-matrix} hold. Then, the estimates\/ $\hat{\xlift}(t),\lambdalift(t),$ and\/ $\thetalift(t)$ generated by \eqref{eq:total-framework} satisfy $\hat{\ubar{x}}(t)\leq x(t)\leq \hat{\bar{x}}(t)$, $\ubar{\lambda}(t)\leq \lambda(t)\leq \bar{\lambda}(t)$ and $\thetalb(t)\leq \theta\leq \thetaup(t)$ for all $t\geq0$.
\end{lemma}

}
\subsection{Recursive feasibility}
In this subsection, we analyze the feasibility of the optimization problem \eqref{eq:mpc-opt-adaptive}. \zl{Recall that the terminal set is $\Xflift=\mc X_f \times \mc X_f$ where $\mc X_f$ is box-constrained with upper bound $\xter$}. There is a trade-off in choosing the bound $\xter_i$ between the mainline cells for $i\in[I]$ and on-ramps for $i\in[I+1:2I]$ as shown in the following result:
\begin{proposition}
	\label{prop:feasible-control-non-empty}
	Let $T$ be sufficiently large. Given $\hat{\xlift}, \lambdalift$, and\/ $\thetalift$ such that\/ $\xlb\leq \xup, \lambdalb\leq\lambdaup, \thetalb\leq\thetaup$,\/ $\mathbbm{P}(\hat{\xlift},\lambdalift,\thetalift)$ is feasible if one of the followings holds:
	\begin{enumerate}
		\item  $\xter_i>0$ and $\xter_{I+i}=\infty, i\in[I]$. \label{prop:feasible-control-non-empty-condition-1}
		\item $\thetalb= \thetaup, \lambdaup=\lambdalb$ and $\xter_i\geq x^{\text{up}}_i ,i\in[I], \xter_i\geq0,i\in[I+1\colon 2I]$ with 
  \begin{equation*}
  \begin{split}
  x^{\text{up}}_I & := \xcrit_I \\
  x^{\text{up}}_i & := \min\{\xcrit_i, \ \xunc_{i} +(x^{\text{up}}_{i+1} - \xunc_{i+1})\dfrac{v_{i+1}}{\beta_iv_i}\}, \\ & \hspace{2in} i\in[I-1] 
  \end{split}
  \end{equation*}
		\label{prop:feasible-control-non-empty-condition-2}
	\end{enumerate}
\end{proposition}	
\begin{remark}
The two conditions in Proposition~\ref{prop:feasible-control-non-empty} correspond to the cases where only jam density is known and all parameters are known, respectively. The latter corresponds to a purely output-feedback case.
\end{remark}
\zl{
Proposition~\ref{prop:feasible-control-non-empty} ensures feasibility of  $\mathbbm{P}(\hat{\xlift}(0),\lambdalift(0),\thetalift(0))$ at time $t=0$ under Assumption~\ref{assu:initial-condition}. Then, we show recursive feasibility, i.e., feasibility at $t=0$ implies feasibility for all $t\geq0$:
\begin{proposition}\label{prop:recursive-feasibility}
    Let Assumptions~\ref{assu:initial-condition}--\ref{ass:C-matrix} hold.
	Suppose that given $\lambdalift$ and $\thetalift$,  $\Xflift$ is positively invariant, i.e., for all $\xlift\in\Xflift$, there exists a control $u\in\mc U$ such that $\Flift(\xlift,u,\lambdalift;\thetalift)\in\Xflift$. Then, $\mathbbm{P}(\hat{\xlift}(0),\lambdalift(0),\thetalift(0))$ is feasible $\implies\mathbbm{P}(\hat{\xlift}(t),\lambdalift(t),\thetalift(t))$ is feasible for all $t\geq0$ under \eqref{eq:total-framework}.
\end{proposition}
}
\begin{example}\label{ex:positively-invariant}
	The proof of Proposition~\ref{prop:feasible-control-non-empty} shows that $\Xflift$ is positively invariant under $u=0$ with $\xter_i>0, i\in[I]$ and $ \xter_i=\infty, i\in[I+1,2I]$, or under $u=\lambda$ with $\xter_i=x^{\text{up}}_i,i\in[I]$ and $\xter_i\geq0,i\in[I+1\colon 2I]$ when $\thetaup=\thetalb,\lambdaup=\lambdalb$.
\end{example}
\begin{remark}
	In nominal state-feedback MPC or output-feedback MPC with observer dynamics, the state values $\hat{\xlift}(t)$ input to the MPC optimization \eqref{eq:mpc-opt-adaptive} come from trajectories of dynamical systems described by difference equations, see \cite{rawlings2017model}. This is not the case for\/ $\hat{\xlift}(t)$ generated by \eqref{eq:total-framework} due to the complications in\/ $\Ulift^x$ and\/ $\Ulift^\theta$. Still, we show recursive feasibility by leveraging monotonicity of the mapping $\Flift$.
\end{remark}
\subsection{Stability}
Now we find conditions under which the closed-loop system consisting of \eqref{eq:cell-mass-balance}, \eqref{eq:output_measurements} in feedback with the Set-PC controller in \eqref{eq:total-framework} is stable. We first show a weaker stability result in terms of the \emph{boundedness} of states when $\theta$ is unknown. Let $l\in\real_{> 0}^{2I}$ be any positive coefficient vector.

\begin{theorem}\label{theo:stability-boundedness}
	Let Assumptions~\ref{assu:initial-condition}--\ref{ass:C-matrix} hold.
	Let\/ $V_f\equiv0$ and\/ $\zl{\ell}(x,u)  = 0$ for all $x\in\mc X_f$, and $\zl{\ell}(x,u)=l^Tx$ for all\/ $x\notin\mc X_f$.
	Let\/ $T$ be sufficiently large and condition~\ref{prop:feasible-control-non-empty-condition-1} in Proposition~\ref{prop:feasible-control-non-empty} hold.
	If the initial condition satisfies\/ $x(0)\in\mc X\setminus\mc X_f$, then there exists a finite time\/ $K$ such that the closed-loop system $\eqref{eq:cell-mass-balance}, \eqref{eq:output_measurements}$ under the Set-PC control \eqref{eq:total-framework} satisfies\/ $x(K)\in\mc X_f$.

	Additionally, if Assumption~\ref{ass:demand-average} holds, \/ $\lambda(t)\equiv\lambda$, and\/ $\xter_i\leq\xunc_i(\lambda),i\in[I]$, then the controller\/ $u(t) = y^r(t) - y^r(t-1) + u(t-1)$
	stabilizes the system \eqref{eq:cell-mass-balance} in the sense that\/ $||x(t)||<\infty$ for all\/ $t\geq0$ and\/ $x(0)\in\mc X_f$.
\end{theorem}


\begin{remark}
	Although the parameter $\xunc(\lambda)$ is unknown, we can use the initial estimates $\lambdalift(0), \thetalift(0)$ to determine $\xter$, e.g.,
		$\xter_i= [(\mathbf{I} - \ubar{\mathbf{R}}(0))^{-1}\ubar{\lambda}(0)]_i/\bar{v}_i(0)\leq\xunc_i(\lambda), i\in[I]$.
\end{remark}
\begin{remark}\label{remark:local-controller}
	Introducing a local controller in Theorem~\ref{theo:stability-boundedness} is motivated by the dual-mode MPC, e.g., see \cite{scokaert1999suboptimal,lazar2008input}. The dual-mode MPC first steers the system trajectory into the terminal set in finite time and then employs a controller that can locally stabilize the system starting from initial conditions within the terminal set.
	The local controller satisfies $u(t) = y^r(t) - y^r(t-1) + u(t-1)=\lambda(t-1)$. An alternative is to average the past observed demand by setting $u(t)=\frac{1}{L-1}\sum_{k=t-L+1}^{t}y^r(k)-y^r(k-1)+u(k)$, which can deal with time-varying demand shown in simulations. Theorem~\ref{theo:stability-boundedness} under the constant demand $\lambda(t)\equiv\lambda$ also holds for this alternative.
\end{remark}

When $\theta$ is known, we show a stronger result in terms of \emph{input-to-state} stability where the exogenous inflow $\lambda(t)$ is interpreted as the \emph{input}. Since commonly used cost functions are linear (c.f. Remark~\ref{remark:cost-function}), we assume $\ell(x,u) = l^Tx$ and $V_f(x,u) = b^Tx$ with $l\in\real_{> 0}^{2I}$ and  $b\in\real_{\geq 0}^{2I}$. We omit the dependence on the parameter $\theta$ in the following result:
\begin{theorem}\label{theo:ISpS}
	Let Assumptions~\ref{ass:demand-average}--\ref{ass:C-matrix} hold,\/ $\lambda(t)\equiv \lambda$\@,\/ $\ell(x,u) = l^Tx$,\/ $V_f(x) = b^Tx$,\/ $T$\@ be sufficiently large, and condition~\ref{prop:feasible-control-non-empty-condition-2} in Proposition~\ref{prop:feasible-control-non-empty} holds. \zl{If there exists\/ $d\in\real_{\geq 0}^I$ such that for all\/ $\xlift\in\Xflift$, there exists a control $u\in \mc U$:
	\begin{equation}\label{eq:terminal-lyapunov}
				b^TF(\xlift,u,\lambdalift) -b^T\bar{x} \leq - l^T\bar{x}+d^T\lambda \\
	\end{equation}
	and $\Flift(\xlift,u,\lambdalift)\in\Xflift$,}
	then the closed-loop system \eqref{eq:cell-mass-balance}, \eqref{eq:output_measurements} under the Set-PC controller \eqref{eq:total-framework} is \emph{input-to-state stable} with respect to\/ $\lambda$, i.e.,
	\begin{equation}\label{eq:ISS-bound}
		\begin{aligned}
	     &\lVert x(t)\rVert_1 \leq \dfrac{\tilde{a}_2}{\tilde{a}_1} \rho^{t}\lVert \bar{x}(0)\rVert_1 + \dfrac{\tilde{a}_3 + (1-\rho)\tilde{a}_2}{\tilde{a}_1(1-\rho)}\lVert \lambda \rVert_1,\\
	     &\hspace{2in} \forall\, t\geq 0,\quad \forall\, x(0)\in\mc X
		\end{aligned}
	\end{equation}
	where $\tilde{a}_1 := \min_i\{l_i\},\tilde{a}_2 :=(T+1) \max\{ \max_i \,l_i , $ $\max_i \, b_i\},\tilde{a}_3 := \max_i\{d_i\}$, and $\rho := 1-\tilde{a}_1/\tilde{a}_2 \in(0,1)$. 
\end{theorem}
\zl{
\begin{remark}
	A key ingredient in the proofs of  Theorems~\ref{theo:stability-boundedness} and \ref{theo:ISpS} is that the cost function is monotone with respect to each component of $x$ due to the positivity of the states. This allows us to construct Lyapunov-like inequalities for the optimal value function.  
\end{remark}
}
\begin{example}\label{ex:controller-parameter}
A combination of\/ $u$\@ together with\/ $d,l,b,\mc X_f$\@ satisfying \eqref{eq:terminal-lyapunov} is as follows. Let $b = \begin{bmatrix}
	b^{mT} & b^{rT}
\end{bmatrix}^T$ with $b^m,b^r\in\real_{\geq 0}^I$ corresponding to terminal cost coefficients of mainline cells and on-ramps, respectively. Pick\/ $u=\lambda, d=b^m$, let $l,b^r$ be arbitrary, $b^m_i\geq[(\IR)^{-1}l]_i/v_i, i\in[I], \xter_i=x^{\text{up}}_i, i\in[I]$ and $\xter_i=0,i\in[I+1\colon 2I]$. Then, for all $\xlift\in\Xflift$, the dynamics $F(\xlift,u,\lambdalift)$ reduce to \eqref{eq:cell_compact_dynamic} with $\foutm_i(\bar{x};\theta) = v_i\bar{x}_i,i\in[I]$, and\/ $\bar{x}_i=0,i\in[I+1\colon 2I]$ since $\xter_i=0$. Therefore, $b^T
	F(\xlift, u,\lambdalift) - b^T\bar{x} = b^{mT} \left(\lambda - (\IR) \foutm_i(\bar{x};\theta) \right) \leq -l^T \bar{x} + b^{mT}\lambda$, where the inequality follows from $b^m_i \geq[(\IR)^{-1}l]_i/v_i$, $i\in[I]$.
\end{example}
\subsection{Parameter estimation}
\label{sec:parameter-estimation}
The Set-PC controller also facilitates parameter estimation. By the Proof of Theorem~\ref{theo:stability-boundedness}, the closed-loop system enters $\mc X_f$ after finite time $K$ and the mainline state stays uncongested thereafter under the control $u(t)=y^r(t)-y^r(t-1)+u(t-1)=\lambda$. For a given $i\in[I]$, we now show that $v_i$ and $\beta_{i-1}$ (if $i>1$) can be uniquely determined if (i) measurements include 
$x_j(t),x_j(t+1),x_j(t+2)$ for all $j\in[i]$; and (ii) $x_j(t)$ lies outside a certain set of measure zero. The condition (i) is naturally a condition on the matrix $\mathbf{C}$ in \eqref{eq:output_measurements} and backward horizon $L$; and (ii) is not practically constraining.

 \begin{figure*}[hbt]
    \centering
    \begin{subfigure}{0.5\textwidth}
		\centering
		\includegraphics[scale = 0.4]{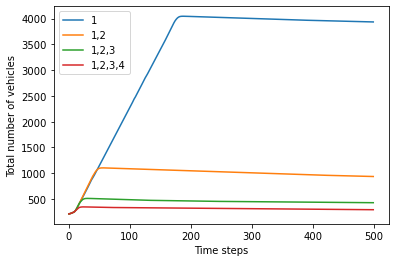}
		\caption{\sf $T=300, L = 1$ with constant demand}
		\label{fig:Set-adaptive-downstream-400-40}
    \end{subfigure}\hfill
    \begin{subfigure}{0.5\textwidth}
		\centering
		\includegraphics[scale = 0.4]{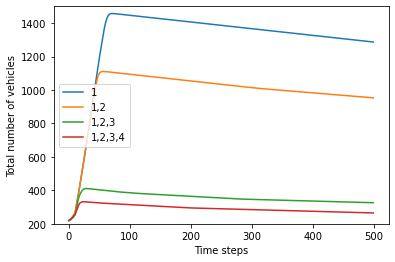}
		\subcaption{\sf $T=300, L=5$ with constant demand}
		\label{fig:Set-adaptive-downstream-300-5}
    \end{subfigure}\hfill
    \begin{subfigure}{0.5\textwidth}
		\centering
		\includegraphics[scale = 0.4]{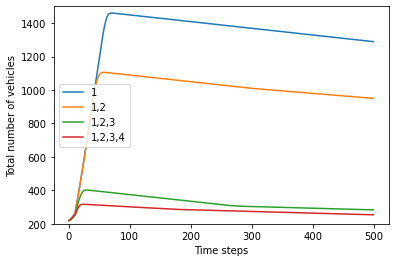}
		\caption{\sf $T=400, L = 5$ with constant demand}
		\label{fig:Set-adaptive-downstream-400-5}
    \end{subfigure}\hfill
    \begin{subfigure}{0.5\textwidth}
		\centering
		\includegraphics[scale = 0.4]{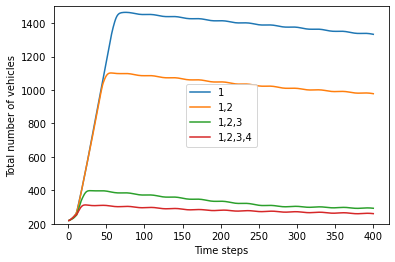}
		\caption{\sf $T=400, L = 5$ with periodic demand  }
		\label{fig:set-adaptive-downstream-periodic}
    \end{subfigure}\hfill
    \caption{\sf Traffic state evolution under the Set-PC controller, for unknown model parameters. The numbering in the legends are the indices of the cells from which measurements are available.}
    \label{fig:set-adaptive-performance}
\end{figure*}
For all $t\geq K$, the dynamics become
\begin{equation}\label{eq:cell-free-flow}
	\begin{aligned}
		x_i(t + 1) &= x_i(t) - v_ix_i(t) + \lambda_i,\quad i=1 \\
		x_i(t+1) &= x_i(t) + \beta_{i-1}v_{i-1}x_{i-1}(t) - v_ix_i(t) + \lambda_i \\ &\hspace{2in} i\in[2:I]
	\end{aligned}
\end{equation}
We first note that $x_i(t)>0$ for all $t>0$ under \eqref{eq:cell-free-flow} if $\lambda_i>0$. This is because $v_i\leq1$ for all $i\in[I]$ (c.f. Remark~\ref{remark:CFL}).
For $i=1$, the only unknown in \eqref{eq:cell-free-flow} is $v_1$, and thus $v_1$ can always be uniquely determined. For $i=2$, we have
\begin{align*}
	x_2(t+2) &\!=\! x_2(t+1) + \beta_{1}v_{1}x_{1}(t+1) - v_2x_2(t+1) + \lambda_2 \\
	x_2(t+1) &\!=\! x_2(t) + \beta_1 v_{1} x_{1}(t) - v_2 x_2(t) + \lambda_2
\end{align*}
The two unknowns are $\beta_1$ and $v_2$. Rearranging the equations gives
\begin{multline}\label{eq:parameter-matrix-form}
	\begin{bmatrix}
		v_1x_1(t+1) & -x_2(t+1) \\
		v_1x_1(t) & -x_2(t)
	\end{bmatrix}
	\begin{bmatrix}
		\beta_1 \\ v_2
	\end{bmatrix} \\ = 
	\begin{bmatrix}
		x_2(t+2) - x_2(t+1) - \lambda_2 \\ x_2(t+1) - x_2(t) - \lambda_2
	\end{bmatrix}
\end{multline}
\zl{Now we want to find conditions under which $\beta_1$ and $v_2$ can be uniquely determined.}
Since $v_1$ is unique, the uniqueness of $\beta_1$ and $v_2$ is equivalent to whether the $2\times 2$ matrix in \eqref{eq:parameter-matrix-form} has full rank, \zl{i.e., whether the determinant is zero. The condition such that the determinant is zero is the following:}
\begin{equation}\label{eq:parameter-rank-condition}
    \frac{x_1(t+1)}{x_1(t)} = \frac{x_2(t+1)}{x_2(t)}
\end{equation}
Substituting \eqref{eq:cell-free-flow} into both sides of \eqref{eq:parameter-rank-condition}
\begin{align*}
    \frac{x_1(t+1)}{x_1(t)} &= 1 - v_1 + \frac{\lambda_1}{x_1(t)} \\
	\frac{x_2(t+1)}{x_2(t)} &= 1 - v_2 + \beta_1v_1\frac{x_1(t)}{x_2(t)} +  \frac{\lambda_2}{x_2(t)}
\end{align*}
Omitting the dependence on $t$ for simplicity, \eqref{eq:parameter-rank-condition} is equivalent to
\begin{multline}\label{eq:parameter-full-rank-x2}
	  \beta_{i-1}v_{i-1}x_{i-1}^2 + (v_{i-1} - v_i)x_{i-1}x_i \\+ \lambda_ix_{i-1} - \lambda_{i-1} x_i= 0,\ i=2
\end{multline} 
Repeating the steps for all $i\in[3:I]$, we have the following condition similar to \eqref{eq:parameter-full-rank-x2}
\begin{multline}\label{eq:parameter-full-rank-xj}
		\beta_{i-1}v_{i-1}x_{i-1}^2 + (v_{i-1} - v_i)x_{i-1}x_i - \beta_{i-2}v_{i-2}x_{i-2}x_i \\+ \lambda_ix_{i-1} - \lambda_{i-1} x_i = 0,\ i\in[3:I]
\end{multline}
Therefore, for each $i\in[2:I]$, $v_i$ and $\beta_{i-1}$ cannot be uniquely determined if there exists $j\in[i]$ such that \eqref{eq:parameter-full-rank-x2} or \eqref{eq:parameter-full-rank-xj} holds. For each $j\in[i]$, the left-hand side of \eqref{eq:parameter-full-rank-x2} or \eqref{eq:parameter-full-rank-xj} is an analytic function that is not identically zero since $x_j=0$ is excluded. Therefore, the set that \eqref{eq:parameter-full-rank-x2} or \eqref{eq:parameter-full-rank-xj} holds for some $j$ is a countable union of the zero sets of an analytic function, which has measure zero by \cite{mityagin2015zero}.





\section{Simulations}
\label{sec:simulations}
\begin{figure*}[hbt]
	\centering
	\begin{subfigure}{0.33\textwidth}
	\centering
	\includegraphics[scale = 0.4]{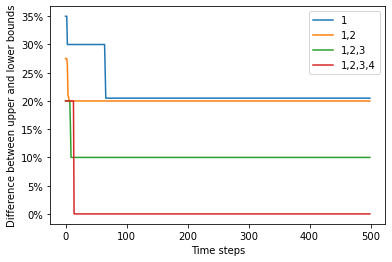}
	\subcaption{\sf Free-flow speed $v$ }
	\label{fig:convergence-v-300-5}
	\end{subfigure}\hfill
	\begin{subfigure}{0.33\textwidth}
	\centering
	\includegraphics[scale = 0.4]{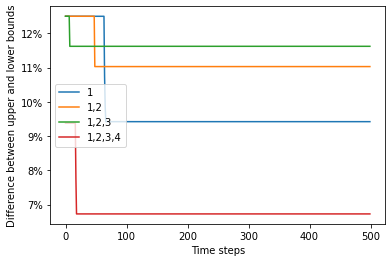}
	\caption{\sf Jam density $\xjam$}
	\label{fig:convergence-jam}
	\end{subfigure}\hfill
	\begin{subfigure}{0.33\textwidth}
	\centering
	\includegraphics[scale = 0.4]{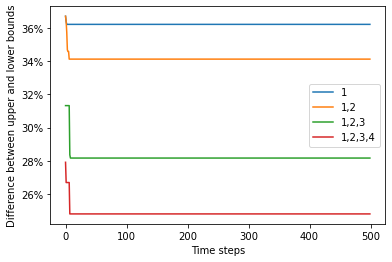}
	\caption{\sf Capacity $\Cm$  }
	\label{fig:convergence-Cm}
	\end{subfigure}\hfill
	\caption{\sf Parameter estimation for $T=300$ and $L=5$}
	\label{fig:performance-parameter-speed}
\end{figure*}
\subsection{Output-feedback adaptive control with unknown $\theta$}
\label{sec:simulation-unknown}
\zl{
\subsubsection{Simulation setups}
}
We first illustrate Theorem~\ref{theo:stability-boundedness}, which applies when $\theta$ is unknown. We conduct simulations on a \zl{freeway stretch} with four mainline cells and four on-ramps. We assume homogeneous cells with parameters  in \cite{gomes2008behavior}: for all $i\in[4]$, \zl{free-flow speed} $v_i=0.5\ \text{cell/time step}$, \zl{congestion wave speed} $w_i=0.5/3$ cell/time step, \zl{jam density} $\xjam_i =160\text{ veh/cell}$, \zl{capacity} $\Cm_i = 20 \text{ veh/time step}$, \zl{turning ratio} $\beta_{i} = 0.9$. The rate of capacity drop is set to be $\alpha_i = 0.9,i\in[4]$.
\zl{We implement the optimization for parameter estimation in \eqref{eq:U-theta-max} and the MPC optimization
\eqref{eq:mpc-opt-adaptive} in \texttt{Gurobi}, which automatically reformulates them as MIBLP and MILP, respectively (cf. Remarks~\ref{remark:mixed-integer-bilinear} and \ref{remark:mixed-integer-linear}}).

We consider two types \zl{of} demand: a constant demand $\lambda(t)\equiv \lambda = [19.17, 1.67, 1.67, 1.67]$ and a periodic demand $\lambda_i(t) = \lambda_i + 0.05\lambda_i\sin(\omega t),i\in[I]$ where $\omega=0.2$. The peak value of the periodic demand violates \eqref{eq:max-flow}, but the demand on average satisfies Assumption~\ref{ass:demand-average}. We set the upper bound $\lambdaup(0)$ to be 10 percent higher than $\lambda$ and the lower bound $\lambdalb(0)$ to be 10 percent lower than $\lambda$.

We consider an initial condition with mainline downstream congestion and zero ramp queue length, i.e., $x_i(0)=30, i\in[3], x_4(0)=120$, and $x_{I+i}(0) = 0, i\in[4]$. We set the initial parameter estimates to be \zl{free-flow speed}: $\bar{v}_i(0) = 0.6, \ubar{v}_i(0) = 0.4$; \zl{congestion wave speed}: $\bar{w}_i(0)=0.3, \ubar{w}_i(0) = 0.1$; \zl{jam density}: $\ubar{x}^{\text{jam}}_i(0)=150, \bar{x}^{\text{jam}}_i(0) = 170$; \zl{capacity:} $ \ubar{C}^{\max}_i=16, \bar{C}^{\max}_i=24$; \zl{turning ratio:} $\ubar{\beta}_i=0.7, \bar{\beta}_i=0.95$ for all $i\in[4]$. We set the initial estimates for mainline states to be $\xup_i(0) = \bar{x}^{\text{jam}}_i(0), \xlb_i(0) = 0,i\in[4]$. Although the theoretical results in Section~\ref{sec:stability-output-mpc} assume that $\ubar{x}^{\text{jam}}(0)=\bar{x}^{\text{jam}}(0)=\xjam$, we relax this assumption for simulations; we only use the upper bound $\bar{x}^{\text{jam}}(t)$ in solving \eqref{eq:mpc-opt-adaptive} by setting $\ubar{x}^{\text{jam}}(t) = \bar{x}^{\text{jam}}(t)$. The closed-loop system might be unstable for some initial condition since $\ubar{x}^{\text{jam}}(t)\geq\xjam$ violates Assumption~\ref{assu:initial-condition}. We provide the following scheme to prevent potential unstable issues. Following the Proof of Theorem~\ref{theo:stability-boundedness}, after $\xup(t)$ enters $\mc X_f$, the controller $u(t) = y^r(t)-y^r(t-1)+u(t-1)=\lambda$ is implemented. In that case, if $x_i(t+1)>x_i(t)$ is observed for some $i\in[I+1,2I]$, then we gradually decrease the value of $\bar{x}^{\text{jam}}(0)$ and restart the closed-loop system under the Set-PC controller \eqref{eq:total-framework}. Repeating the steps makes $\bar{x}^{\text{jam}}(t)$ close to $\xjam$ and the resulting closed-loop system stable. \zl{Whether there exists an initial condition that leads to instability is not clear since Assumption~\ref{assu:initial-condition} is only sufficient but not necessary. In our simulations, we did not find an initial condition that leads to instability}, and thus such an iteration was not used.

\begin{figure}[h]
	\centering
	\includegraphics[width=0.7\linewidth]{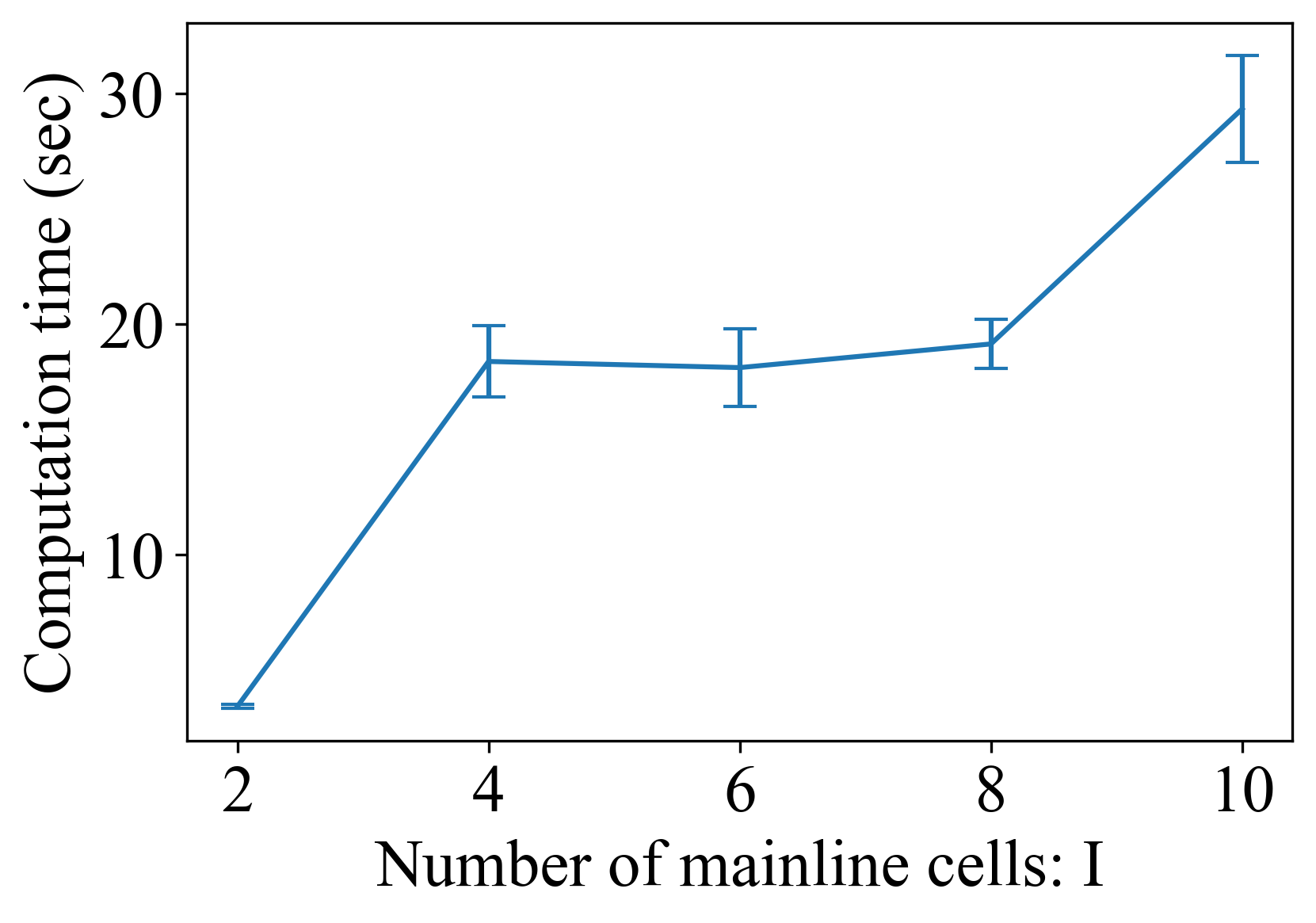}
	\caption{\sf Computation time of \eqref{eq:mpc-opt-adaptive} under different $I$}
	\label{fig:computation-time}
\end{figure}

\zl{
\subsubsection{Performance evaluation}
}
For controller parameters, we set the coefficient $l$ to be an all-one vector.
When only the state of the first cell is measured, i.e., the matrix $\mathbf{C}$ in \eqref{eq:output_measurements} is such that $c_{1,1}\neq0, c_{i,i}=0,i\in[2:4]$, we obtain a lower bound for prediction horizon to be $T\geq215$ based on the proof of Proposition~\ref{prop:feasible-control-non-empty}. Therefore, we choose $T$ to be 300 and 400 to evaluate the effect of $T$. Following Remark~\ref{remark:local-controller}, for all $i\in[I]$ we use the controller $u_i(t)=\epsilon_i + \frac{1}{L-1}\sum_{k=t-L+1}^{t}y^r_i(k)-y^r_i(k-1)+u_i(k-1)$ with $\epsilon_i=0.1$ for both constant and periodic demands after $x(t)\in\mc X_f$ until $x_{I+i}(t)=0$ for any $i\in[I]$. The additional $\epsilon_i$ is introduced to decrease the ramp queue length and does not violate stability if it is sufficiently small. When only the state of the first cell is measured, the average computation time for \eqref{eq:mpc-opt-adaptive} is 209.6 seconds with $T=300$ and 324.6 seconds with $T=400$. The average computation time for \eqref{eq:U-theta-max} is 0.9 seconds with $L=1$ and 4.1 seconds with $L=5$. We also observed that further increasing the value of $L$ significantly increases the computation time for \eqref{eq:U-theta-max}. 
\zl{Then, we set the solver to terminate after the optimality gap of \eqref{eq:mpc-opt-adaptive}, i.e., the difference between upper and lower bounds of objective value divided by the upper bound, is below $1\%$ and evaluate how computation time scales with different values of $I$. Fig.~\ref{fig:computation-time} 
shows that the computation time remains similar when $I$ increases from $4$ to $8$.
}
These computation times are for a desktop with an Intel Core i7 2.1 GHz Processor and 16 GB RAM. 

\begin{figure*}[!htb]
	\centering
	\begin{subfigure}{0.33\textwidth}
		\centering
		\includegraphics[scale = 0.4]{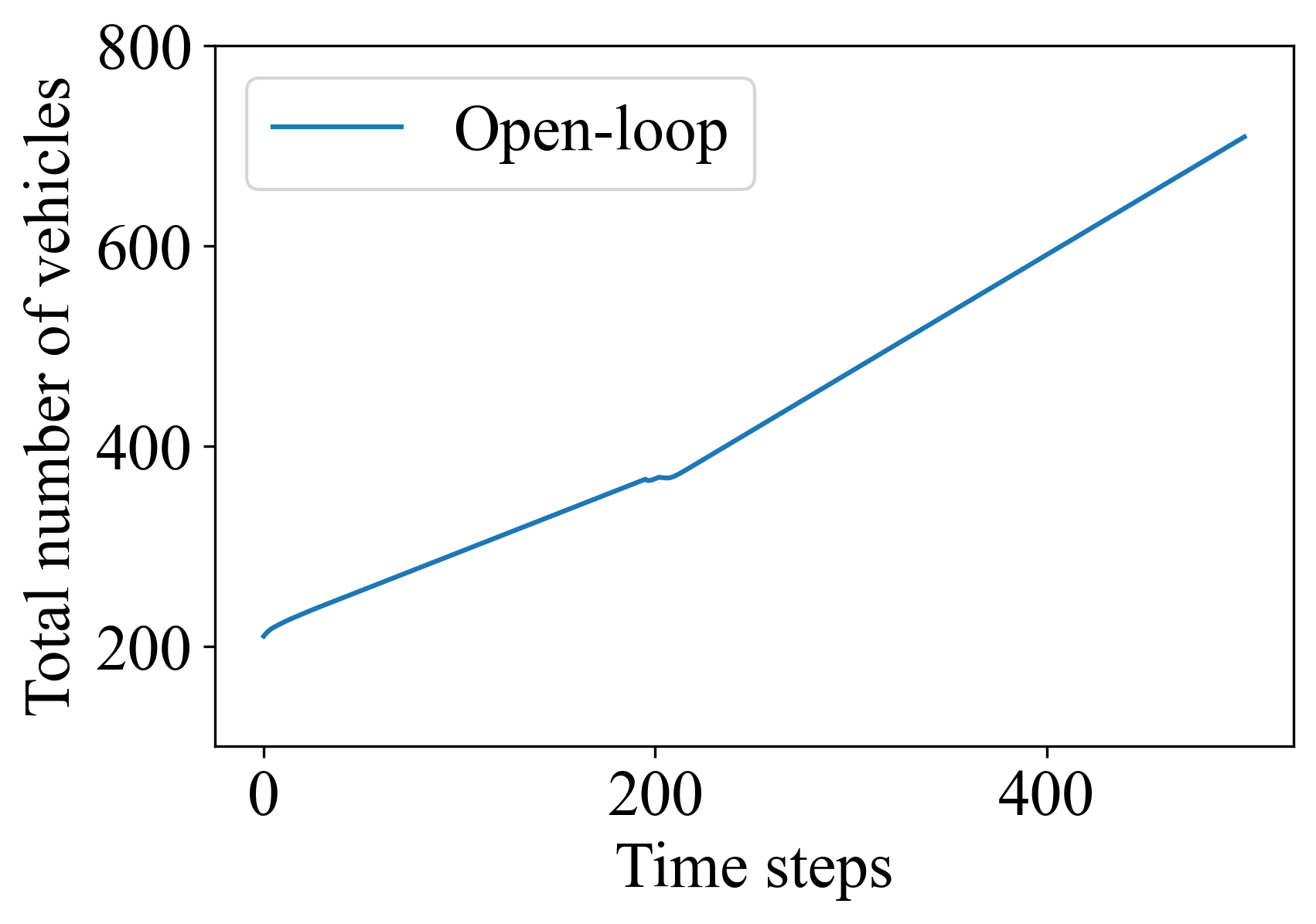}
		\caption{\sf Open-loop system with $u(t)=\lambda$ }
		\label{fig:open-loop-down}
	\end{subfigure}\hfill
	\begin{subfigure}{0.33\textwidth}
		\centering
		\includegraphics[scale = 0.4]{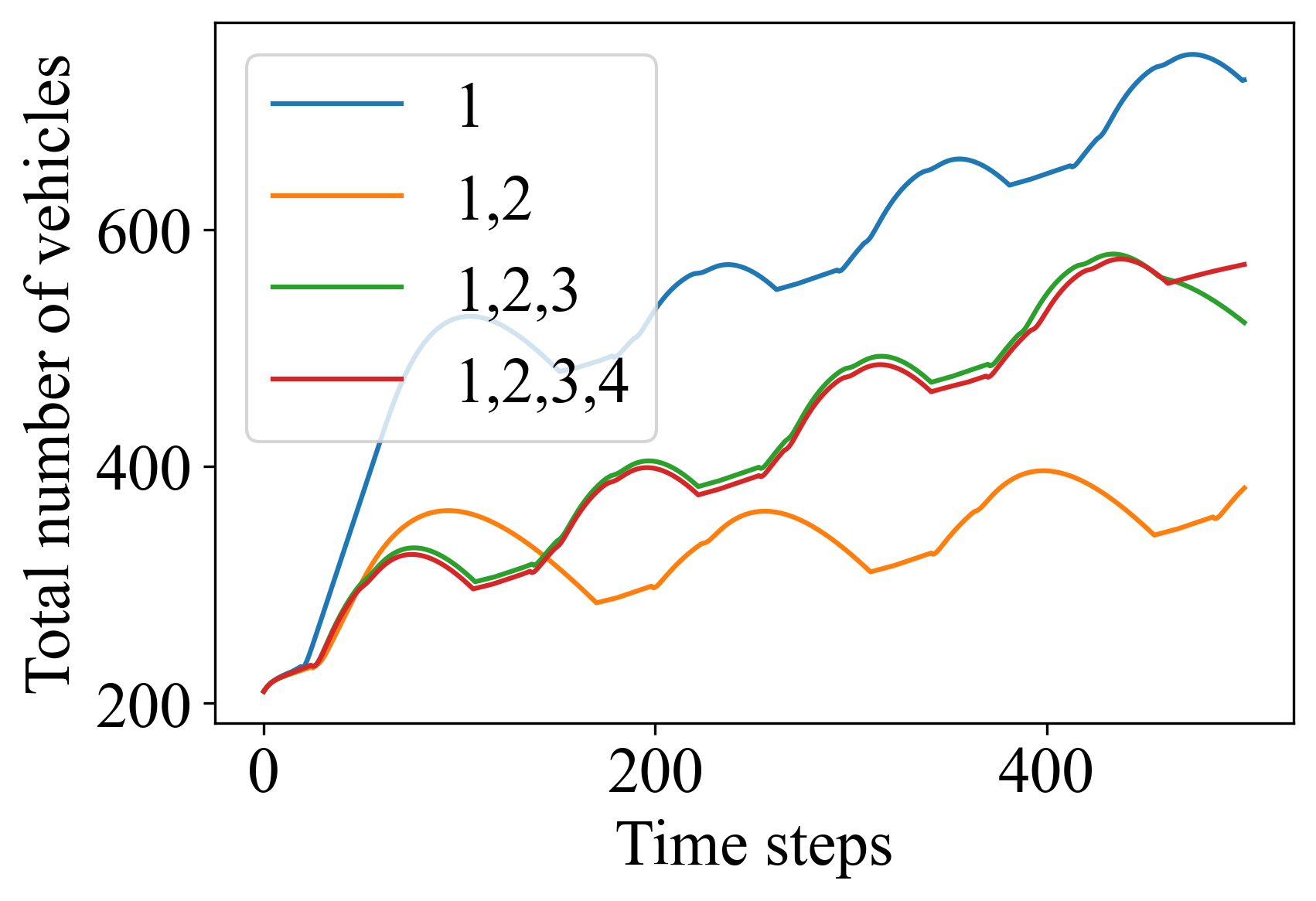}
		\caption{\sf ALINEA }
		\label{fig:ALINEA-down}
	\end{subfigure}\hfill
	\begin{subfigure}{0.33\textwidth}
		\centering
		\includegraphics[scale = 0.4]{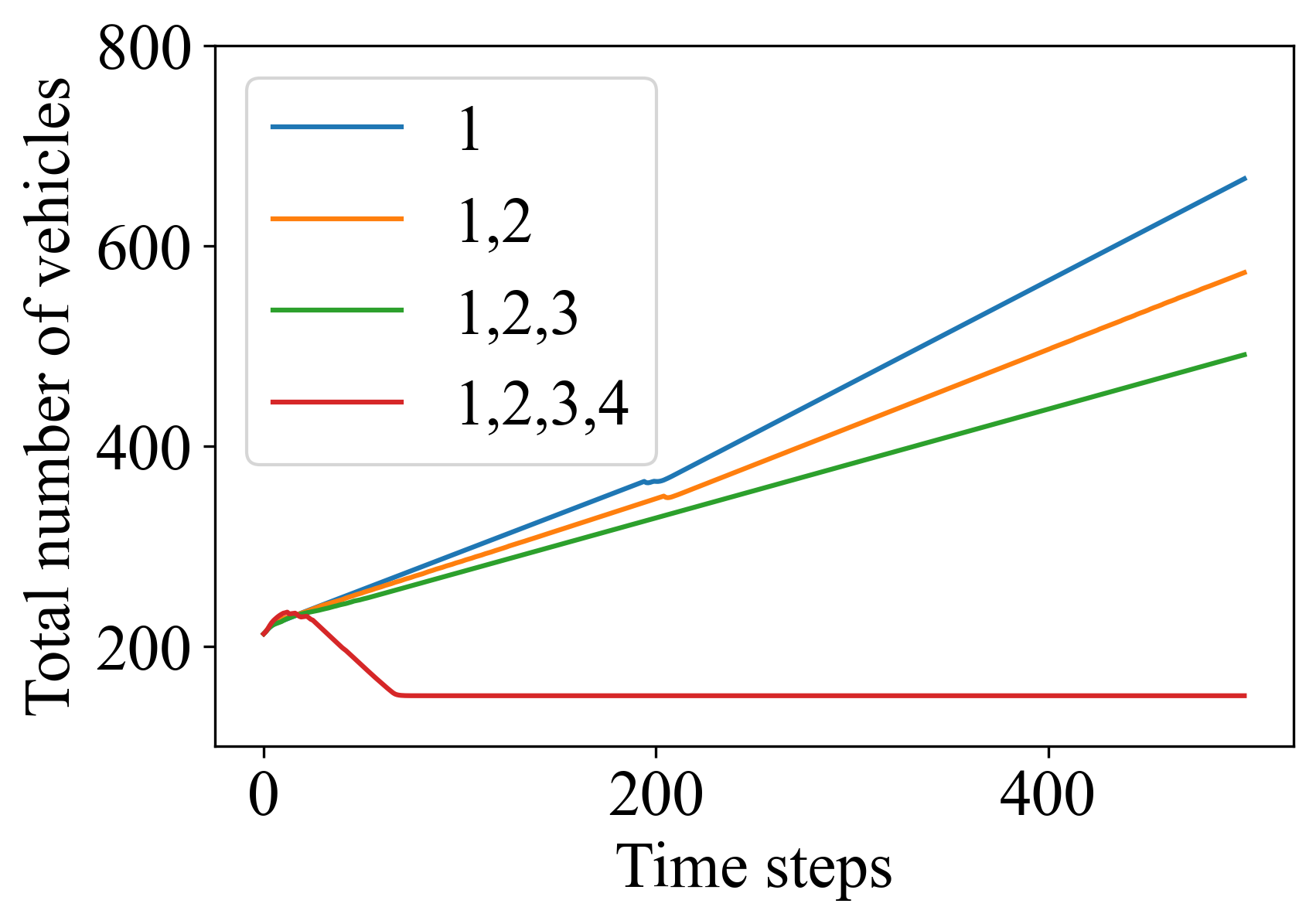}
		\caption{\sf MPC-Luenberger }
		\label{fig:Luenberger-down}
	\end{subfigure}\hfill
	\begin{subfigure}{0.33\textwidth}
		\centering
		\includegraphics[scale = 0.4]{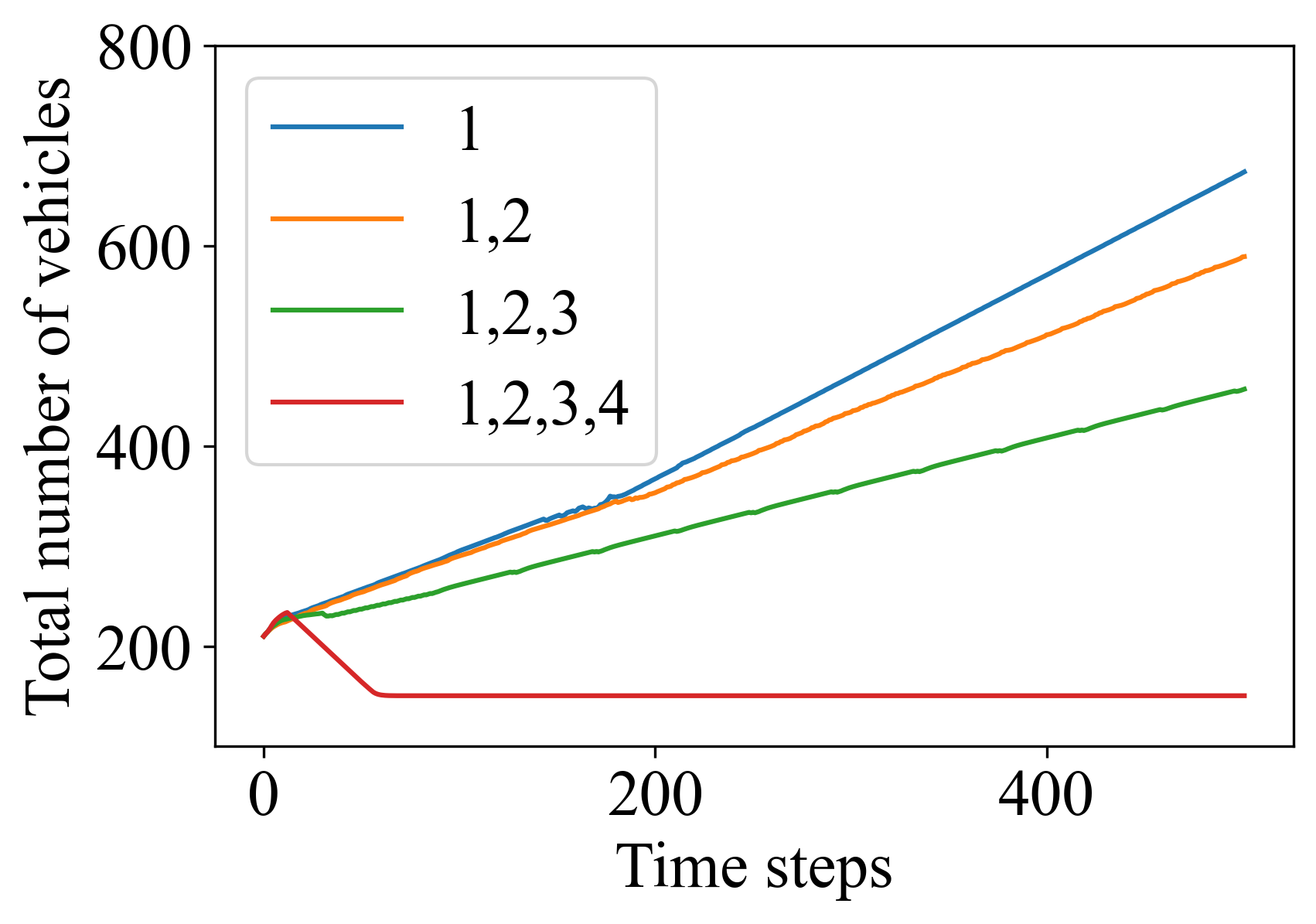}
		\caption{\sf Sequential MPC-MHE }
		\label{fig:sequential-mpc-mhe-down}
	\end{subfigure}\hfill
	\begin{subfigure}{0.33\textwidth}
		\centering
		\includegraphics[scale = 0.4]{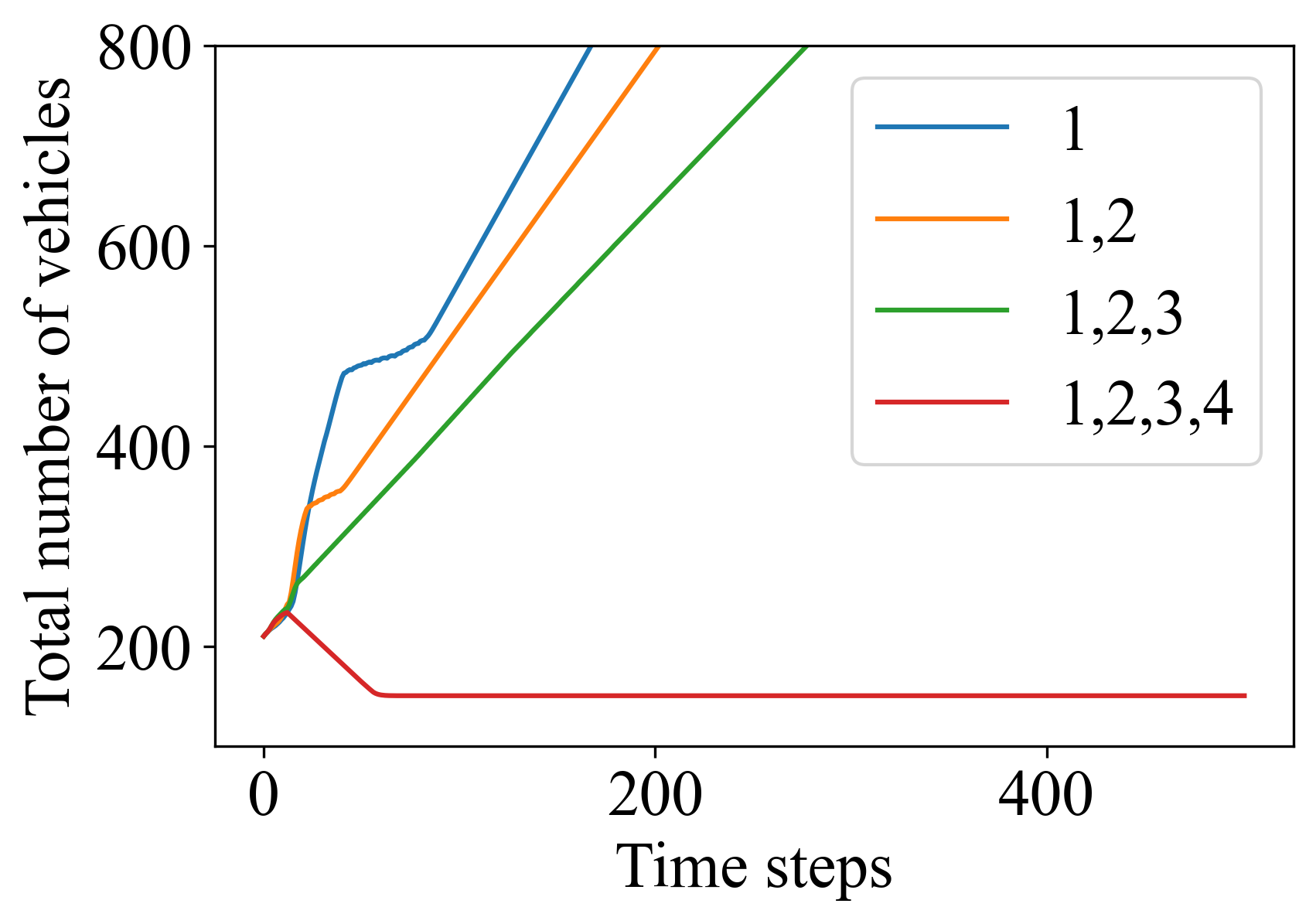}
		\caption{\sf Joint MPC-MHE}
		\label{fig:joint-mpc-mhe-down}
	\end{subfigure}\hfill
	\begin{subfigure}{0.33\textwidth}
		\centering
		\includegraphics[scale = 0.4]{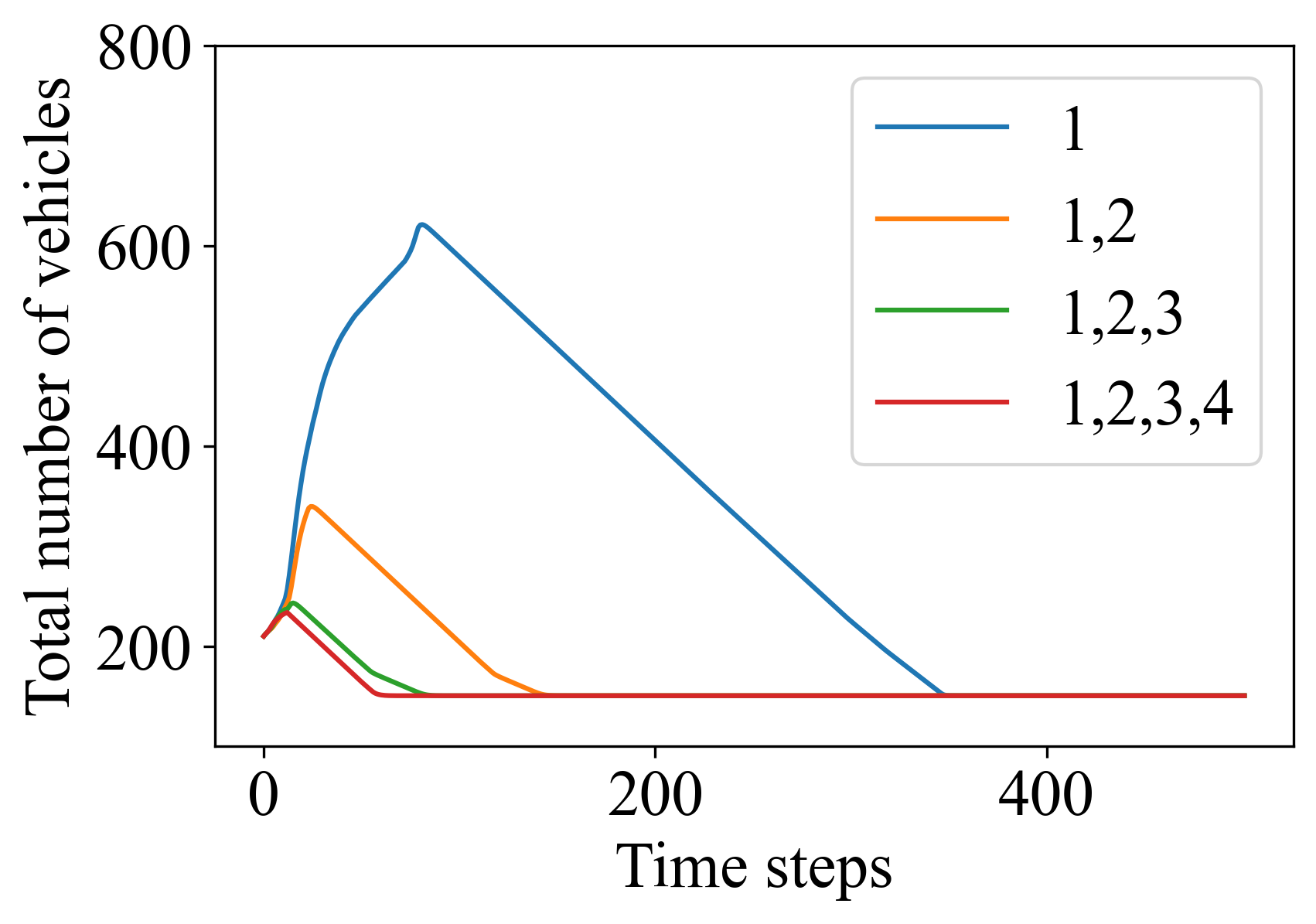}
		\caption{\sf Set-PC}
		\label{fig:mpc-set-membership-down}
	\end{subfigure}\hfill
	\caption{\sf Traffic state evolution starting from the downstream congested initial condition, under different controllers. The numbering in the legends is the indices of the cells from which measurements are available.}
	\label{fig:downstream-congested}
\end{figure*}
\begin{figure*}[!tb]
	\centering
	\begin{subfigure}{0.33\textwidth}
		\centering
		\includegraphics[scale = 0.4]{"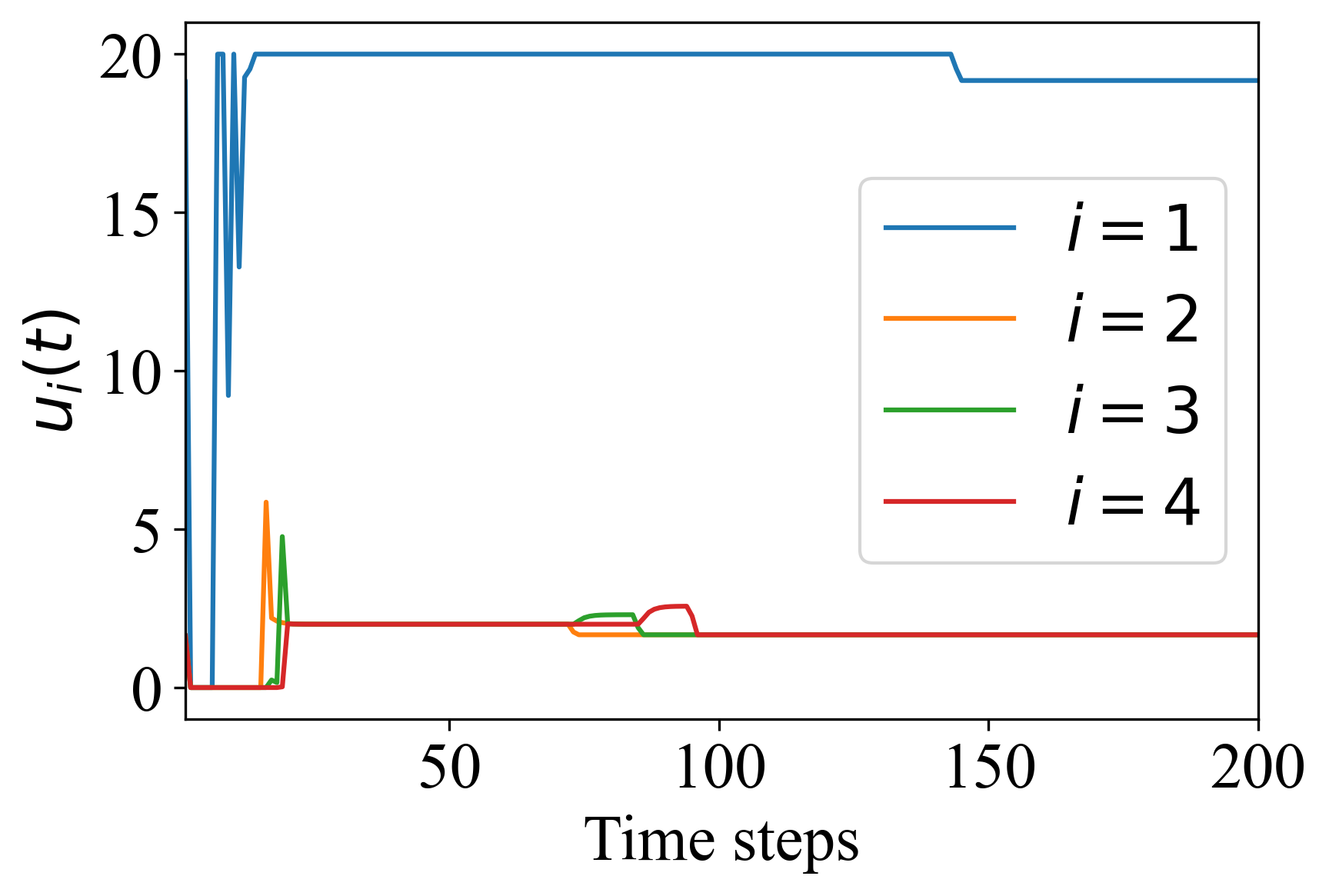"}
		\caption{\sf Set-PC: Ramp metering rate }
		\label{fig:ramp-metering-rate-set}
	\end{subfigure}\hfill
	\begin{subfigure}{0.33\textwidth}
		\centering
		\includegraphics[scale = 0.4]{"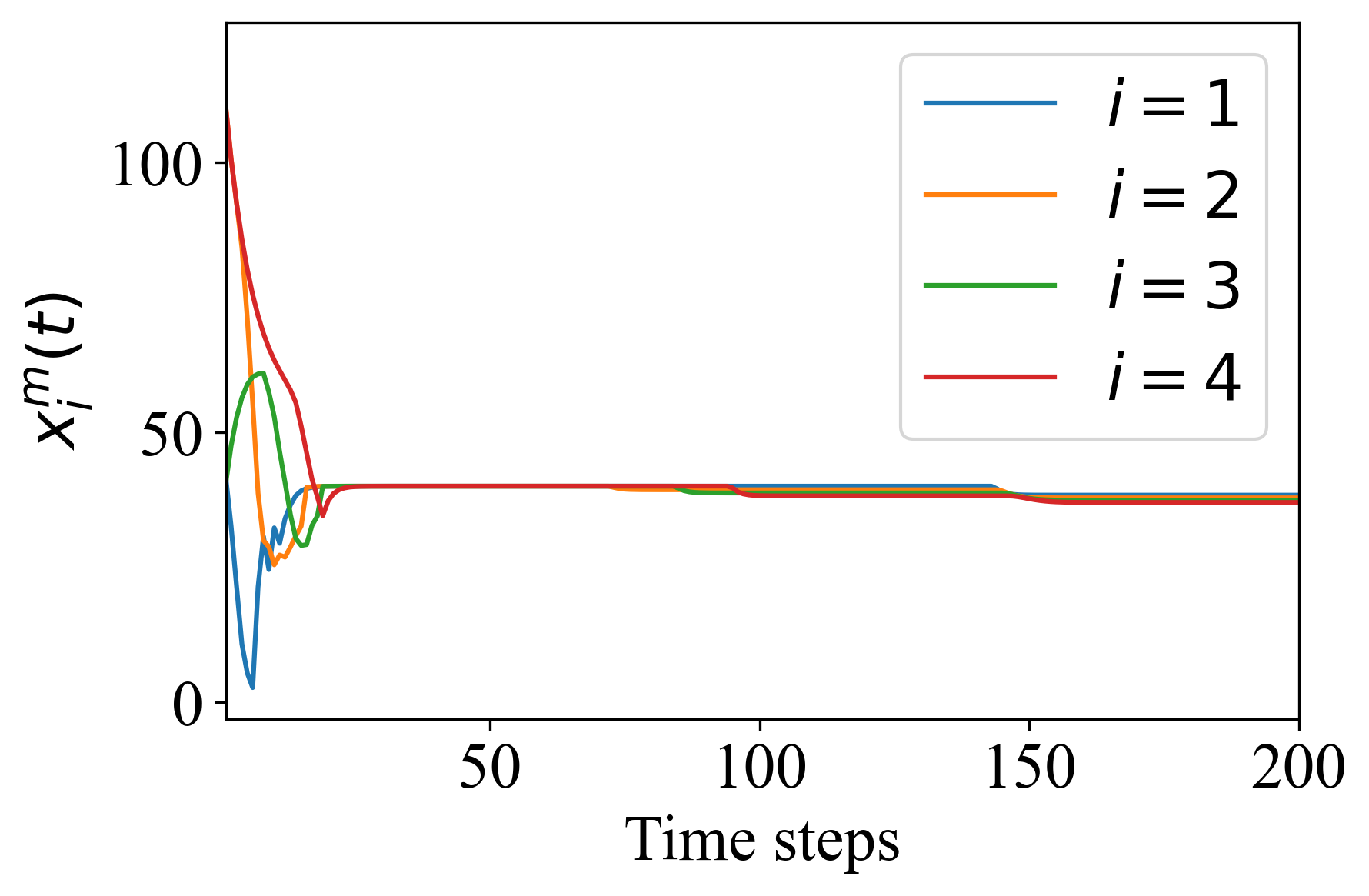"}
		\caption{\sf Set-PC: Mainline state }
		\label{fig:mainline-state-set}
	\end{subfigure}\hfill
	\begin{subfigure}{0.33\textwidth}
		\centering
		\includegraphics[scale = 0.4]{"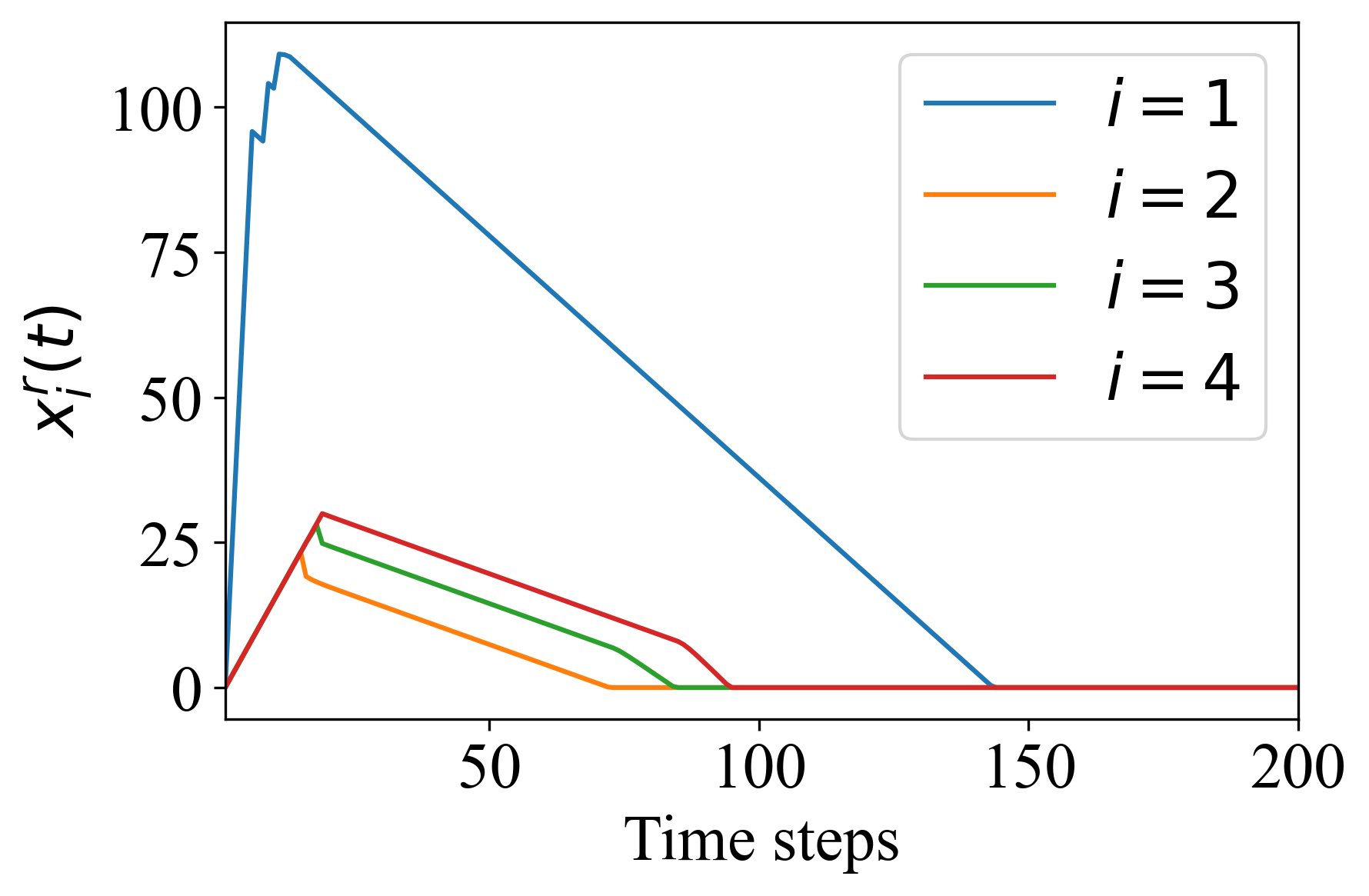"}
		\caption{\sf Set-PC: Ramp state }
		\label{fig:ramp-state-set}
	\end{subfigure}\hfill
	\begin{subfigure}{0.33\textwidth}
		\centering
		\includegraphics[scale = 0.4]{"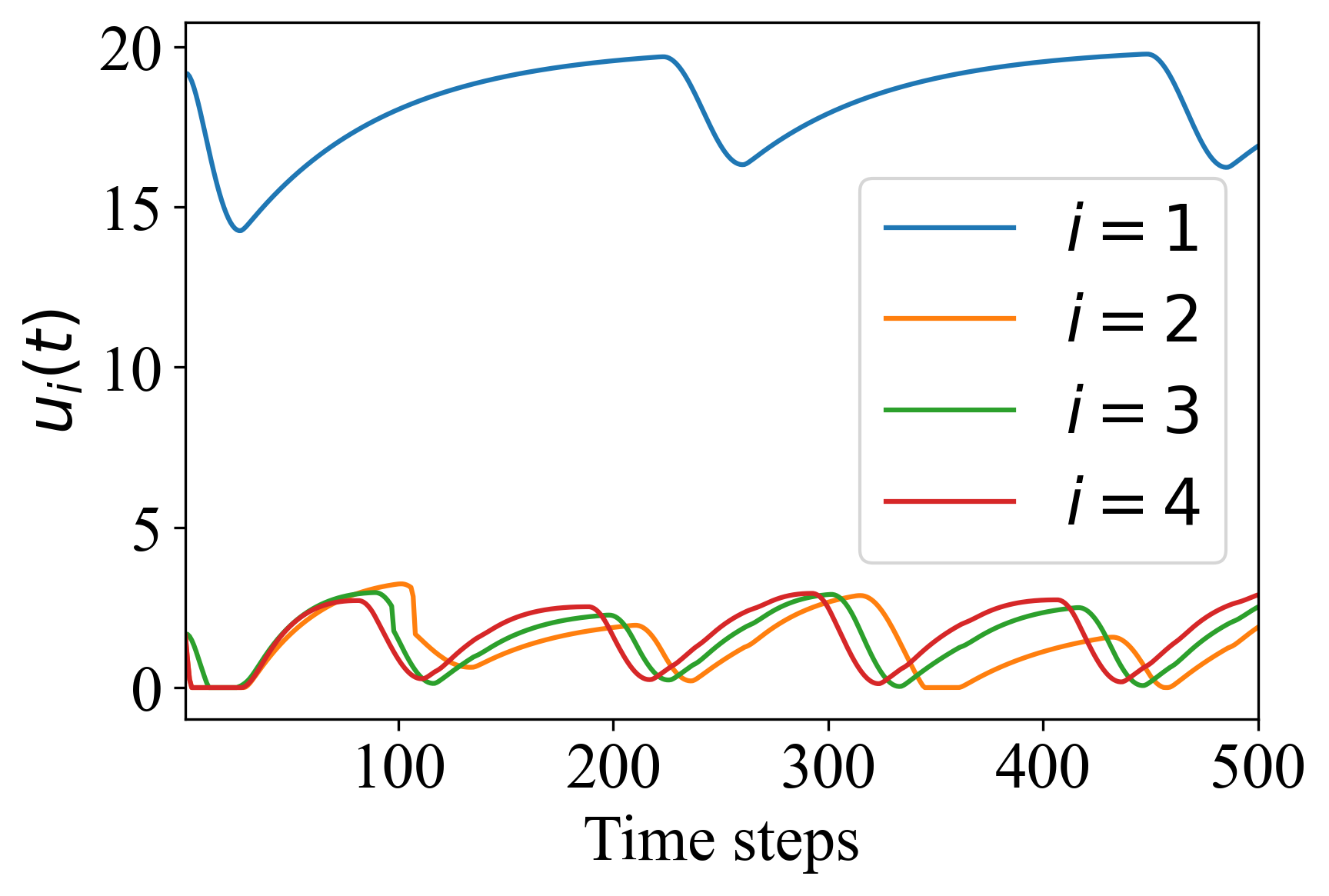"}
		\caption{\sf ALINEA: Ramp metering rate }
		\label{fig:ramp-metering-rate-alinea}
	\end{subfigure}\hfill
	\begin{subfigure}{0.33\textwidth}
		\centering
		\includegraphics[scale = 0.4]{"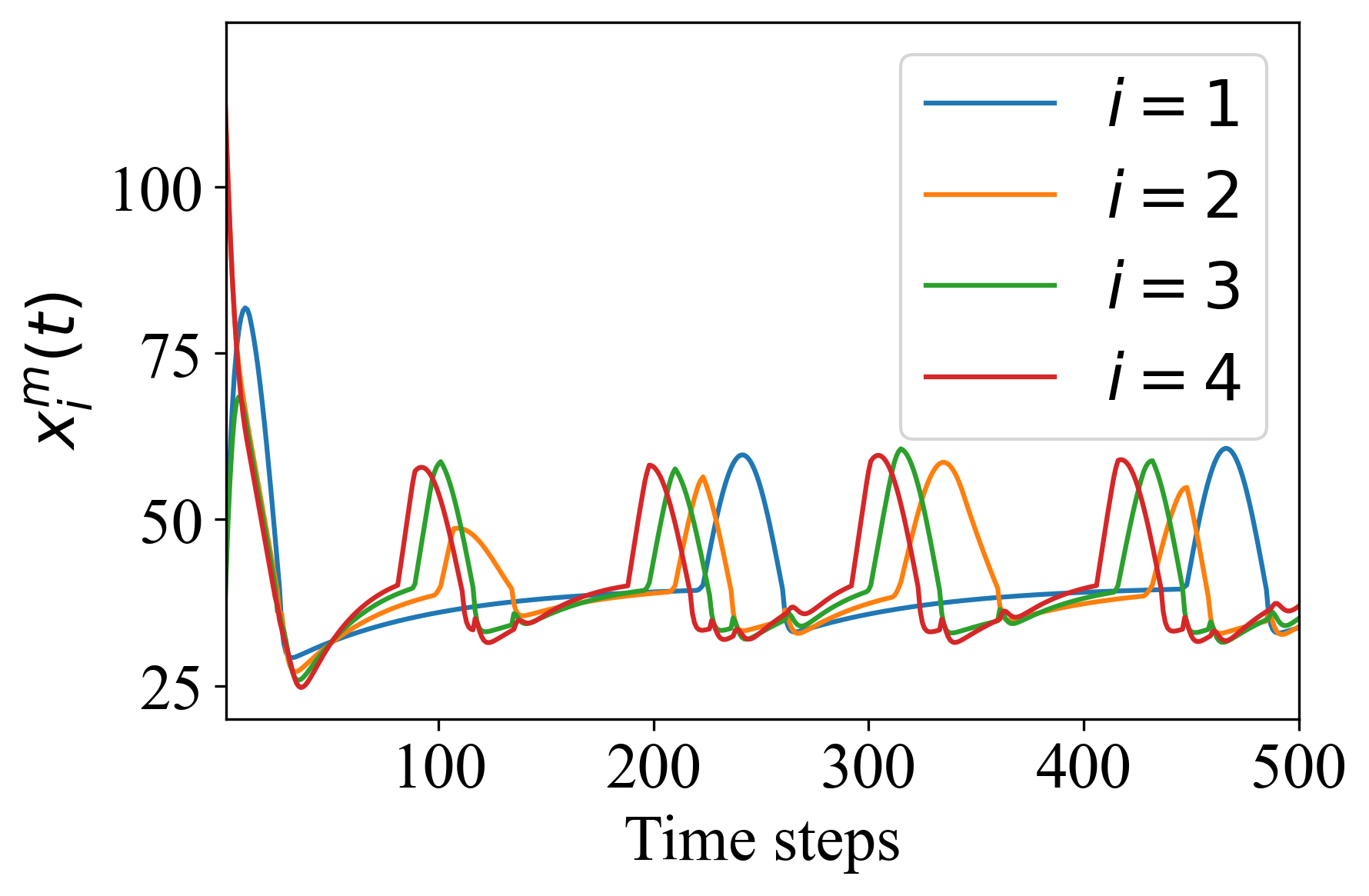"}
		\caption{\sf ALINEA: Mainline state }
		\label{fig:mainline-state-alinea}
	\end{subfigure}\hfill
	\begin{subfigure}{0.33\textwidth}
		\centering
		\includegraphics[scale = 0.4]{"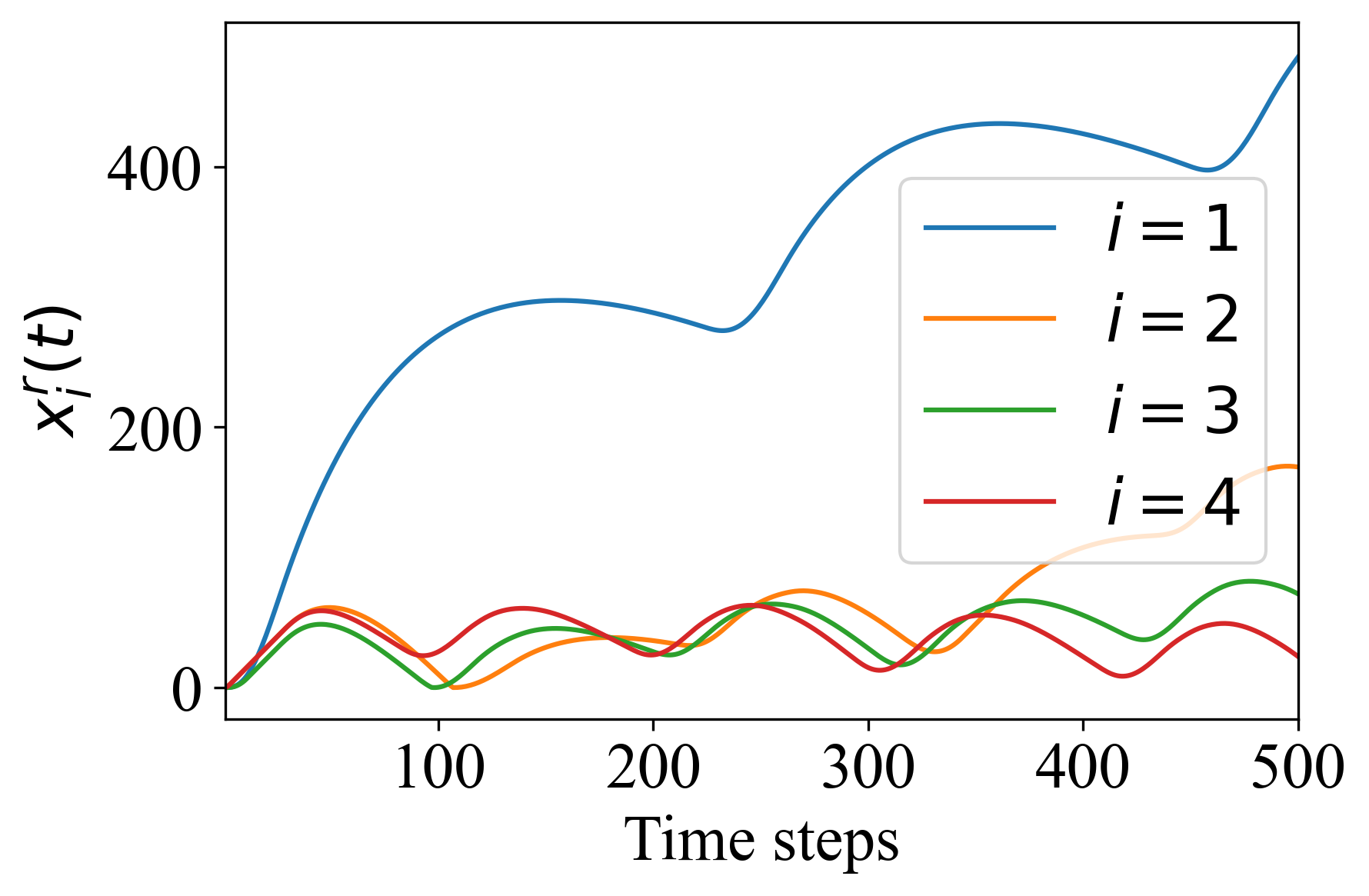"}
		\caption{\sf ALINEA: Ramp state }
		\label{fig:ramp-state-alinea}
	\end{subfigure}\hfill
	\begin{subfigure}{0.45\textwidth}
		\centering
		\includegraphics[scale = 0.4]{"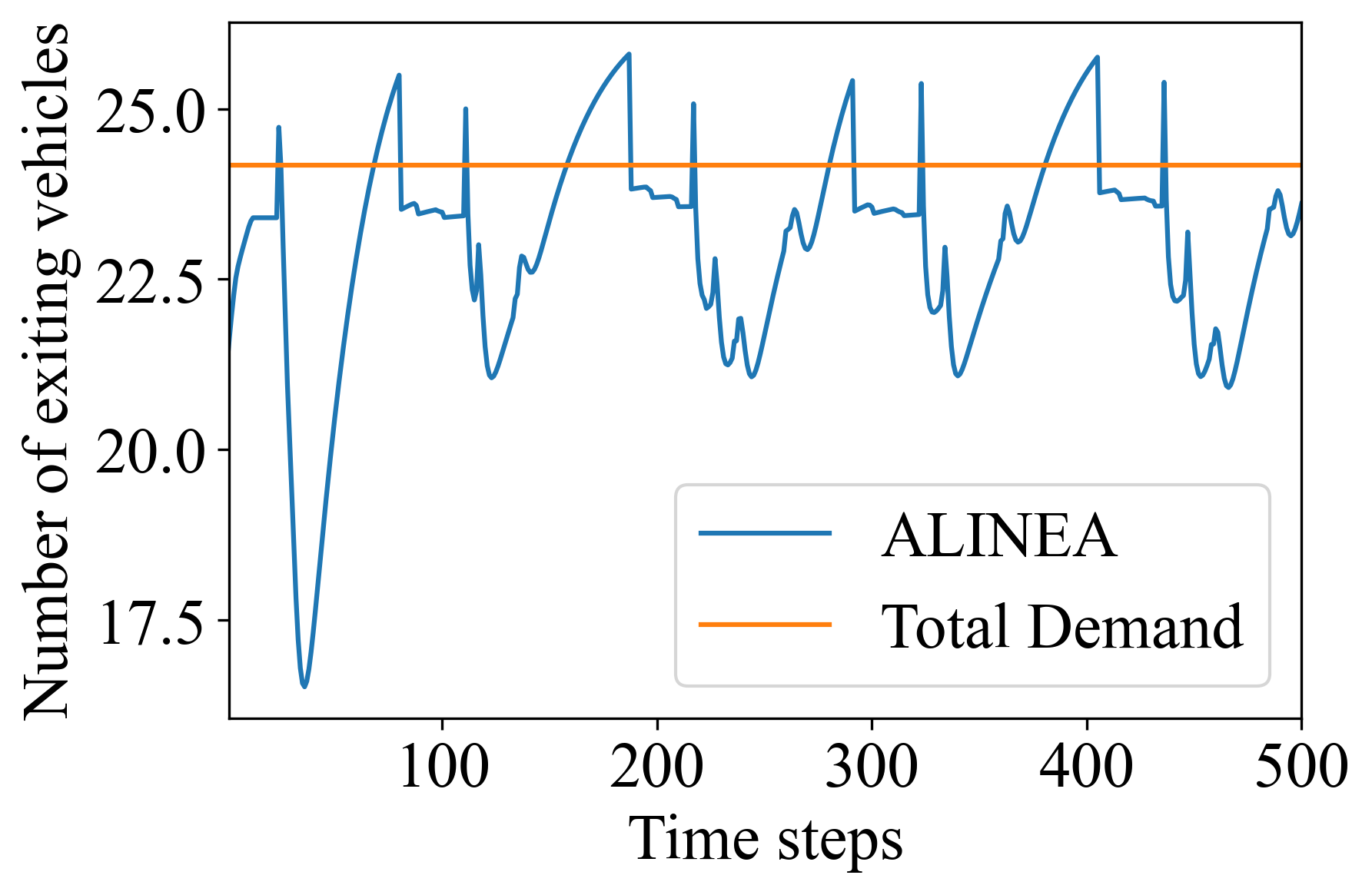"}
		\caption{\sf Throughput under ALINEA }
		\label{fig:outflow-alinea}
	\end{subfigure}\hfill
	\begin{subfigure}{0.45\textwidth}
		\centering
		\includegraphics[scale = 0.4]{"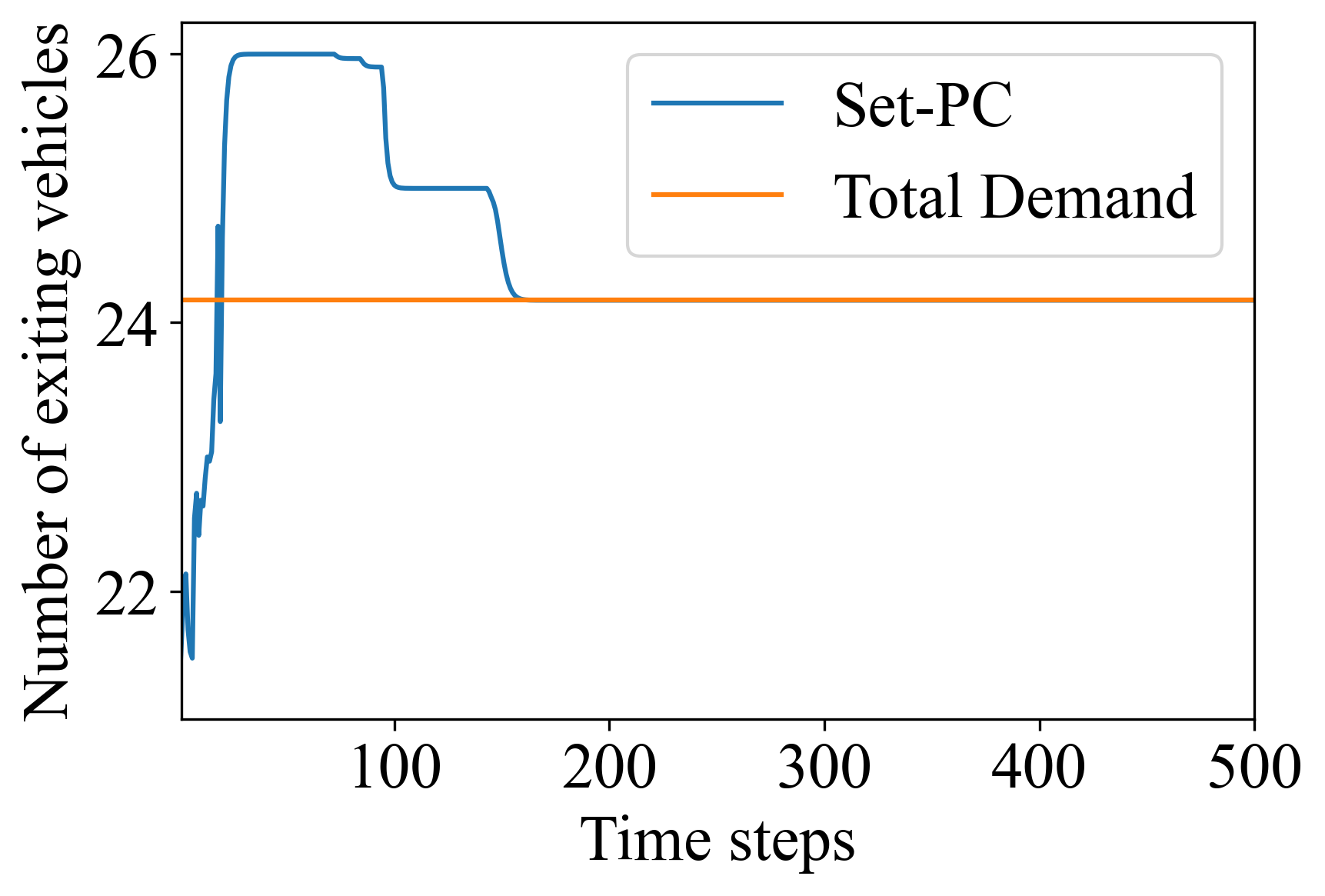"}
		\caption{\sf Throughput under Set-PC }
		\label{fig:outflow-set}
	\end{subfigure}\hfill
	\caption{\sf Comparison of ALINEA and Set-PC when measurements from mainline cells 1 and 3 are available}
	\label{fig:control-performance}
\end{figure*}
\begin{figure*}
	\centering
	\begin{subfigure}{0.5\textwidth}
		\centering
		\includegraphics[scale = 0.4]{"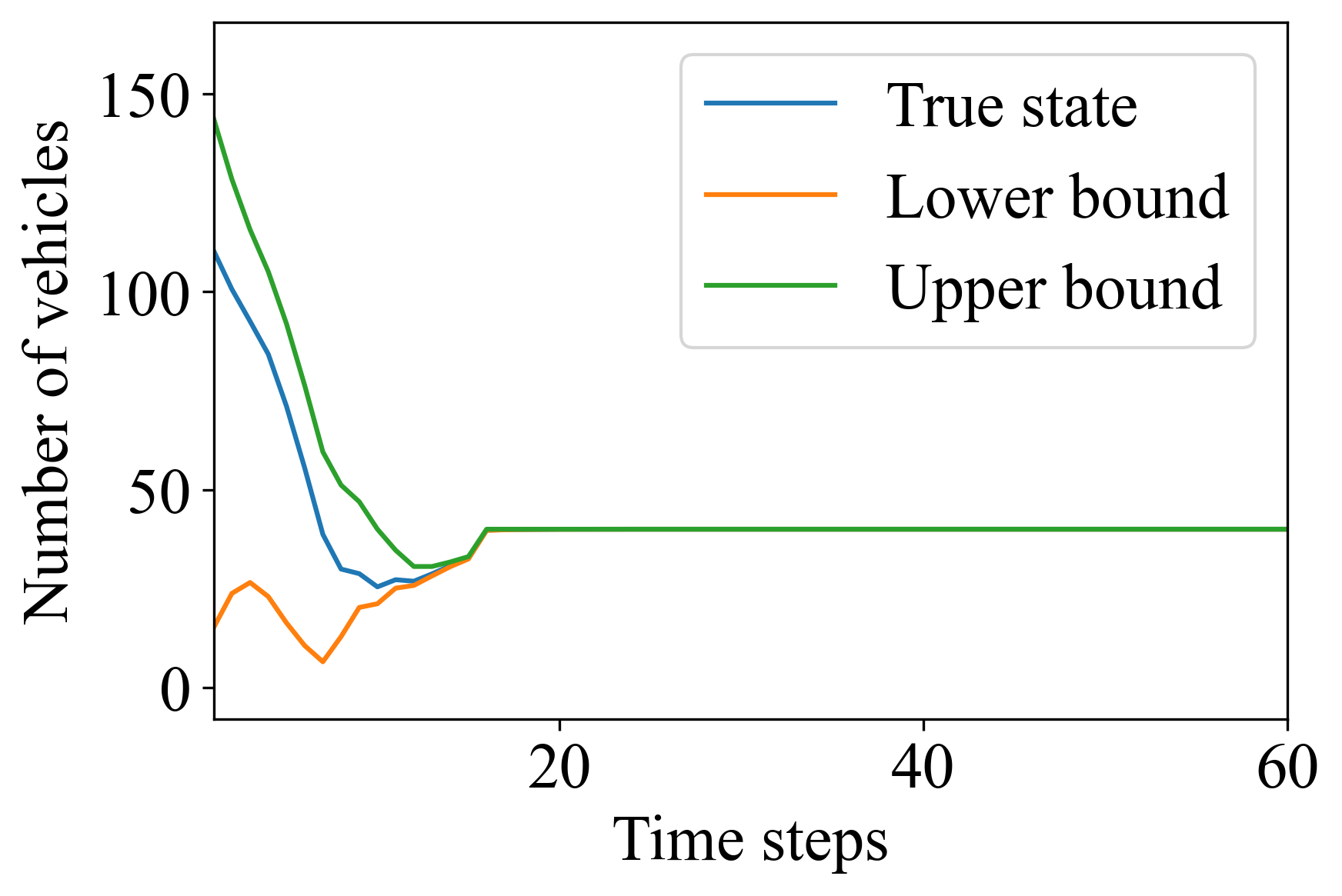"}
		\subcaption{\sf Mainline cell 2 }
		\end{subfigure}\hfill
		\begin{subfigure}{0.5\textwidth}
			\centering
			\includegraphics[scale = 0.4]{"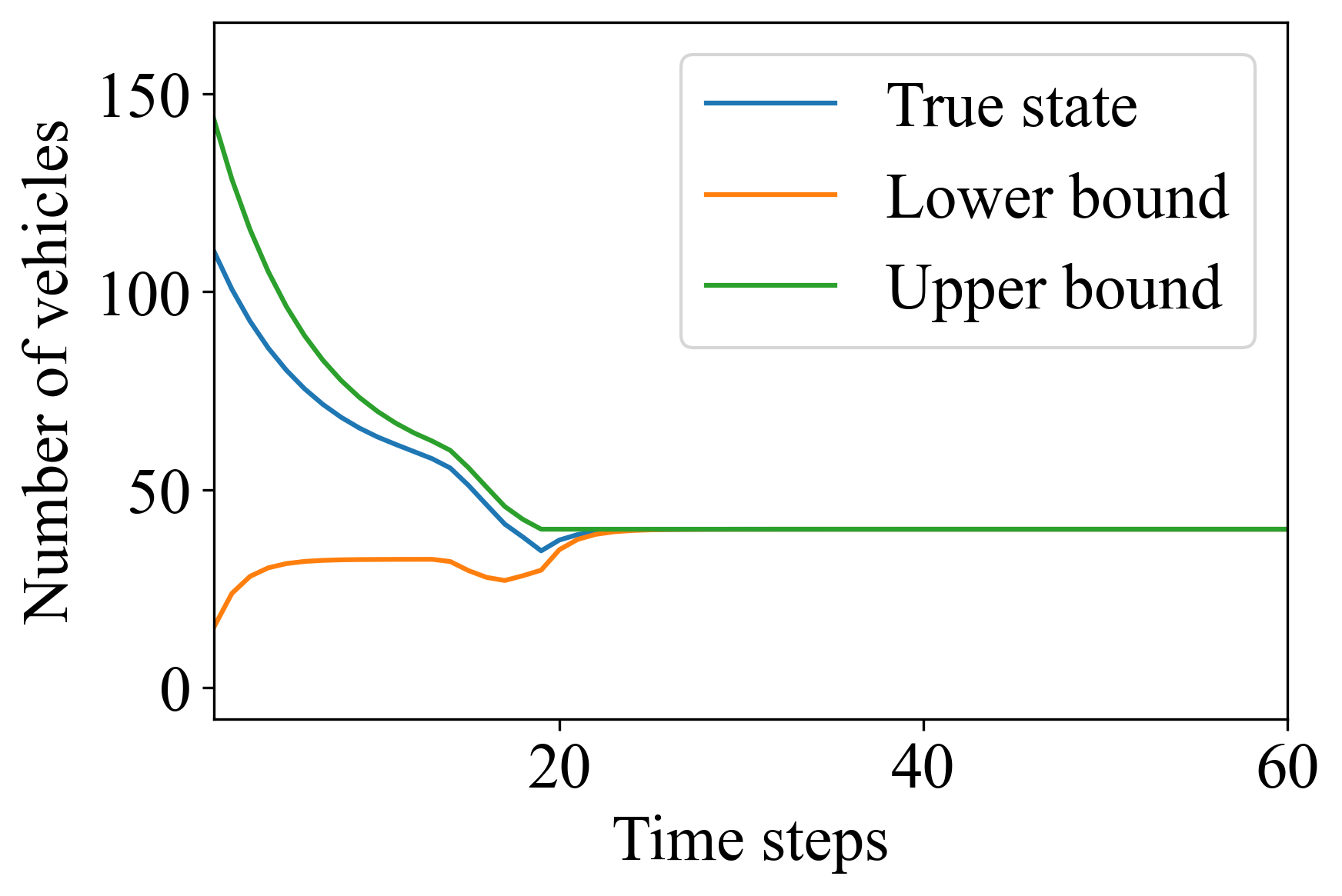"}
			\subcaption{\sf Mainline cell 4}
			\end{subfigure}\hfill
	\caption{\sf Performance of state estimation for mainline cells without measurements}
	\label{fig:performance-state-estimation}
\end{figure*}
We use the total number of vehicles or total queue lengths, i.e., $\| x(t) \|_1=\sum_{i \in [2I]} x_i(t)$, to evaluate the performance of the Set-PC controller.
Figure~\ref{fig:set-adaptive-performance} shows the performance under different forward horizon $T$ and backward horizon $L$ values. It can be seen that the Set-PC controller stabilizes the system under both constant and periodic demands. Increasing the value of $L$ improves the transient performance, but increasing the value of $T$ does not affect the transient performance significantly. 

We then evaluate the performance of parameter estimation in terms of the difference between upper bound $\thetaup(t)$ and lower bound $\thetalb(t)$. We choose free-flow speed $v$, jam density $\xjam$, and capacity $\Cm$ as examples. As shown in Figure~\ref{fig:convergence-v-300-5}, the difference between the upper and lower bounds of $v$ decreases as the number of measured cells increases, and the difference becomes zero when all cells are measured. This is consistent with the derivation in Section~\ref{sec:parameter-estimation}, where we show that the free-flow speed $v$ can be uniquely determined within the uncongested region. As for jam density $\xjam$ and capacity $\Cm$, Figures~\ref{fig:convergence-jam} and \ref{fig:convergence-Cm} show that there remains a gap between the upper and lower bounds, which implies that we are not able to estimate these parameters exactly. 
We can still obtain maximal stability of the closed-loop system because the Set-PC controller first steers the system to the uncongested region, and the parameters we cannot uniquely determine are irrelevant from then on.

\subsection{Output-feedback control with known $\theta$}

When $\theta$ and $\lambda(t)\equiv\lambda$ are known, we compare the performance of the proposed Set-PC controller and output-feedback controllers. The controller setups are summarized in Appendix~\ref{sec:appendix-existing-control}.
For MPC-based controllers, cost coefficients $l$ and $b^r$ are all-one vectors, and $b^m$ is chosen according to Example~\ref{ex:controller-parameter}.

\zl{Figure~\ref{fig:downstream-congested} illustrates stability properties for increasing nonzero entries in $\mathbf{C}$ under different controllers. 
The closed-loop system is unstable if total number of vehicles, i.e., $\|x(t)\|_1$, grows unbounded. 
As shown in Figures~\ref{fig:open-loop-down}--\ref{fig:ALINEA-down}, the open-loop and ALINEA controllers fail to stabilize the system, even with complete state information. This means that the demand exceeds the throughput provided by these two controllers. Figures~\ref{fig:Luenberger-down}--\ref{fig:joint-mpc-mhe-down} show the performance of MPC \eqref{eq:mpc-opt-adaptive} augmented with existing estimation methods instead of set-membership estimation. 
It can be seen that they stabilize the system only with complete state information, meaning throughput decreases with incomplete state information.
On the contrary, the Set-PC controller stabilizes the system with incomplete state information, as shown in Figure~\ref{fig:mpc-set-membership-down}.
This is consistent with Theorem~\ref{theo:ISpS} that the Set-PC controller renders maximal throughput with incomplete mainline state information.

To understand why Set-PC is stabilizing whereas ALINEA is not, we conduct more simulations with initial conditions $x_1(0)=x_3(0)=30,x_2(0)=x_4(0)=120$, and measurements from mainline cells 1 and 3 are available.
Fig,~\ref{fig:control-performance} shows the performance in terms of state values, metering rates, and throughputs, i.e., number of vehicles existing the freeway. 
From Fig.~\ref{fig:mainline-state-set}, Set-PC quickly steers mainline states to the uncongested region and remains there to reduce ramp queue lengths to zero.
In terms of throughput, Fig.~\ref{fig:outflow-set} shows that the throughput quickly converges to the demand, as expected for stable systems.
On the other hand, Fig.~\ref{fig:mainline-state-alinea} shows that ALINEA does not keep the mainline states uncongested, leading to oscillations between congested and uncongested regions.
Since capacity drop occurs in the congested region, the ramp queue lengths increase, and the throughput decreases, when the mainline is congested as shown in Fig.~\ref{fig:ramp-state-alinea} and Fig.~\ref{fig:outflow-alinea}.

Figure~\ref{fig:performance-state-estimation} shows state estimates for mainline cells 2 and 4 without measurements.
The state estimates successfully serve as the upper and lower bounds of actual states, as expected from Lemma~\ref{lemm:set-membership}.
Also, the state estimates converge to the true states as differences between the upper and lower bounds converge to zero.
}


\section{Conclusion and Future Work}
\label{sec:conclusion}
Ramp metering is an effective method for freeway traffic control. There is practical and methodological interest in developing feedback ramp metering control for incomplete measurements and unknown model parameters. In this paper, we integrate set-membership estimation methodology with the MPC framework to address this challenge. The key enabler for tight stability analysis of the closed-loop system is \zl{positivity} of the state and mixed-monotone embedding for the parameterized traffic flow dynamics, which forms the modeling basis for estimation and control. \zl{In the case of constant demand and unbounded on-ramps}, the resulting closed-loop system is shown to render maximal throughput to a freeway stretch and, in particular, shows superior stability properties to existing ramp metering control methodologies when only a few measurements are available on the mainline.
\zl{The proposed Set-PC controller can also benefit other application domains that exhibit mixed-monotonicity and positivity, such as gas and water distribution networks.}


There are several avenues for further research. 
The set-membership estimator in this paper was designed for asymptotic performance. We plan to investigate generalizations for better transient performance and to enable the estimation of all unknown model parameters. Extending the results to general network configuration, other output models, and general traffic flow models with model uncertainty and measurement noise will be interesting.
\zl{More realistic time-varying demands and scenarios, including bounded on-ramps, will be investigated.}
It is also of interest to explore if the spatial sparsity of the dynamics implies sparsity in the structure of the output feedback controller along the lines of our recent work on MPC-based traffic flow control~\cite{Jafari.Savla:Automatica20}.

\bibliographystyle{ieeetr}
\bibliography{references}

\section{Appendix}
\subsection{Output-feedback controller setups}\label{sec:appendix-existing-control}
\begin{enumerate}
	\item Open-loop: $u(t)=\lambda$ without the use of measurements, which replicates the uncontrolled case.
	\item  ALINEA: A \zl{local} ramp metering algorithm widely used in the real world \cite{papageorgiou1991alinea}. 
	\zl{For each on-ramp $I+i,i\in[I]$: 
	$
	u_i(t) = u_i(t-1) - K_R(\xcrit_i - \hat{\bar{x}}_i(t)),
	$ where the gain $K_R=70/60/160$ per time step where $70$ veh/hr is recommended in \cite{papageorgiou1991alinea}, $60$ time steps/hr means that each time step corresponds to 1 min, and $160$ veh is the jam density used in simulations; $\hat{\bar{x}}(t)$ is the state estimate from \eqref{eq:total-framework}; set-point is set to be $\xcrit$. Recall that $\forall\, i\in[I], \xcrit_i=\Cm_i/v_i=40 \text{ veh}$ is the critical state at which the outflow is equal to capacity.}
 \item MPC-Luenberger: using Luenberger estimation, e.g., see \cite{nugroho2021should,brandi2017model} and solving the MPC problem \eqref{eq:mpc-opt-adaptive} with $x(0|t) = z(0|t)$ so that the embedding dynamics $F$ in \eqref{eq:mixed-monotone-decomposition-function} reduce to the original dynamics \eqref{eq:cell-mass-balance}.
	\item   Sequential model predictive control-moving horizon estimation (MPC-MHE): using optimization-based MHE, e.g., see \cite{sirmatel2019nonlinear}, to compute state estimates and then solve the MPC problem \eqref{eq:mpc-opt-adaptive} with $x(0|t) = z(0|t)$ for the same reason in MPC-Luenberger. The upper and lower bounds generated by \eqref{eq:total-framework} are incorporated as additional constraints for the MHE estimation. 
	\item Joint MPC-MHE: formulating a min-max problem inspired by \cite{copp2017simultaneous} and solving the maximization problem for state estimates and minimization problem for control inputs separately. The upper and lower bounds generated by \eqref{eq:total-framework} are incorporated as additional constraints in the maximization problems. 
\end{enumerate}
\subsection{Technical Results}
In this section, we collect technical results to be used later in the proofs. 

The following \textit{monotone} property of the mainline dynamics in \eqref{eq:cell_compact_dynamic} is proven in \cite[Lemma 5.1]{gomes2008behavior}. 
			\begin{lemma} \label{lemm:monotone_x}
				Consider the dynamics in \eqref{eq:cell_compact_dynamic} with $\xi(x_i;\theta)\equiv \Cm_i,i\in[I]$. For any states $x,\tilde{x}\in\mc X(\theta)$ such that $x\leq \tilde{x}$, the following holds: $$x^{m} - (\IR)\foutm(x;\theta) \leq \tilde{x}^m - (\IR)\foutm(\tilde{x};\theta)$$ 
			\end{lemma}

\subsection{Proof of Proposition~\ref{prop:mixed-monotone}}
\label{proof:mixed-monotone}
The first condition (M1)
follows by $\tilde{d}_i(x_i,x_i;\theta,\theta) = d_i(x_i;\theta), \tilde{s}_i(x_i;\theta,\theta) = s_i(x_i;\theta)$ for all $i\in[I]$. 


For (M2) and (M3), the monotonicity in $\lambda$ is trivial since $\lambda$ only appears in $\Fin_i,i\in[I+1,2I]$. For monotonicity in parameters, we want to show that $F$ is non-increasing in $\eta$ and non-decreasing in $\theta$. For all $i\in[I]$, the function $\Fin_i$ is only non-increasing in $\tilde{v}_i$ of the function $\xi$. Then we put $\tilde{v}_i$ in $\eta$ and all other parameters in $\theta$; the function $-\Fout_i$ is only non-decreasing in $\beta_i$ and $v_i$ of the function $\xi$. Then we put these two parameters in $\theta$ and all other parameters in $\eta$. By \eqref{eq:mixed-monotone-decomposition-function}, $F$ is non-decreasing in $\theta$ and non-increasing in $\eta$.

Next, we omit the dependence of $F$ on demand $\lambda$ and parameters $\theta,\eta$ to simplify notations and show that $F$ is non-decreasing in $x$ and non-increasing in $z$. We start by noting that $\tilde{d}_i$ is non-increasing in $z_i$ for all $i\in[I]$, and thus $F$ is non-increasing in $z$. 
Then, consider any $x, \tilde{x} \in \mc X$ such that $x\leq \tilde{x}$, we show $F_i(x,z)\leq F_i(\tilde{x},z) , i\in [2\colon 2I]$, and $F_1(x,z)\leq F_1(\tilde{x},z)$ follows similarly. Consider one cell at a time:
   \begin{enumerate}[leftmargin=0pt,itemindent=2em]
   \item For $x_{i-1}\leq \tilde{x}_{i-1}, x_i=\tilde{x}_i,x_{i+1}=\tilde{x}_{i+1}, i\in[2\colon I-1] \colon  F_i(x,z)\leq F_i(\tilde{x},z)$ since $ \tilde{d}_{i-1}(x_{i-1}, z_{i-1})\leq \tilde{d}_{i-1}(\tilde{x}_{i-1}, z_{i-1})$.
   \item For $x_{i-1} = \tilde{x}_{i-1}$, and $x_i\leq \tilde{x}_i$ or $x_{i+1}\leq \tilde{x}_{i+1},i\in[2:I-1]$: There exists a $\check{x}_{i-1}$ such that $\tilde{d}_{i-1}(x_{i-1},z_{i-1}) = \tilde{d}_{i-1}(\tilde{x}_{i-1},z_{i-1})  = d_{i-1}(\check{x}_{i-1})$. By Lemma~\ref{lemm:monotone_x} and the fact that $(\check{x}_{i-1},x_i,x_{i+1})\leq (\check{x}_{i-1},\tilde{x}_i,\tilde{x}_{i+1})$ where the inequality is element-wise, we have
	$F_i(x,z) - F_i(\tilde{x},z) = x_i - \tilde{x}_i +  \beta_{i-1}(\min\{d_{i-1}(\check{x}_{i-1}), s_i(x_i)\} $ $-\min\{d_{i-1}(\check{x}_{i-1}), s_i(\tilde{x}_i)\}  ) - \min\{d_i(x_i), s_{i+1}(x_{i+1})\} + \min\{ d_i(\tilde{x}_i), s_{i+1}(\tilde{x}_{i+1})\}  \leq 0$.
    \item For $i=I$, this is a special case of the above when $x_{i+1}$ disappears in the dynamics.
	\item For $i\in[I+1\colon 2I]$, $F_i(x,z)\leq F_i(\tilde{x},z)$ follows by $F_i(x,z)=x_i+\lambda_{i-I}-u_{i-I}$.
   \end{enumerate}

\subsection{Proof of Lemma~\ref{lemm:set-membership}}
\label{proof:set-membership}
The properties of $\lambdalift(t)$ and $\thetalift(t)$ follows from the construction of $\Ulift^\lambda$ and $\Ulift^\theta$ in Section~\ref{sec:state-demand-parameter-estimates}.
Since Assumption~\ref{assu:initial-condition} ensures that $\tilde{\ubar{x}}(0)\leq x(0)\leq \tilde{\bar{x}}(0)$,
we will show that $\tilde{\ubar{x}}(t)\leq x(t)\leq \tilde{\bar{x}}(t)$ implies $\ubar{x}(t)\leq x(t) \leq \bar{x}(t)$ and $\tilde{\ubar{x}}(t+1)\leq x(t+1)\leq \tilde{\bar{x}}(t+1)$ for all $t\geq0$ in the remaining part of the proof.

From Assumption~\ref{ass:C-matrix}, we have $y_i(t)/c_{i,i}=x_i(t)$ if $c_{i,i}\neq0$. Then, by the construction of\/ $\Ulift^x$ in Section~\ref{sec:state-demand-parameter-estimates}, $\tilde{\ubar{x}}(t)\leq x(t)\leq \tilde{\bar{x}}(t)$ implies $\ubar{x}(t)\leq x(t) \leq \bar{x}(t)$.

From Assumption~\ref{assu:initial-condition}, jam density is given and therefore $\Xlift(\thetalift(t)) = \mc X(\theta)\times\mc X(\theta)$. Then, the state constraint $\hat{\xlift}(k|t)\in\Xlift(\thetalift(t))$ in \eqref{eq:mpc-opt-adaptive} ensures that for all $t\geq0$, $\fout_{I+i}(x(t),u(t),\lambda(t);\theta)=u_i(t),i\in[I]$ under \eqref{eq:total-framework}. 
From Proposition~\ref{prop:mixed-monotone}, the embedding dynamics $\Flift$ used in \eqref{eq:predict} satisfy (M1)--(M3). 
The properties (M1)--(M3) imply $\tilde{\ubar{x}}(t+1)\leq x(t+1)\leq \tilde{\bar{x}}(t+1)$ since $\ubar{x}(t)\leq x(t) \leq \bar{x}(t), \thetalb(t)\leq \theta\leq\thetaup(t)$, and $\lambdalb(t)\leq\lambda(t)\leq\lambdaup(t)$.

\subsection{Proof of Proposition~\ref{prop:feasible-control-non-empty}}
\label{proof:prop-feasibility}
We omit the dependence of variables on $t$ to simplify notations.
Case \ref{prop:feasible-control-non-empty-condition-1} holds with the control $\hat{u}(k) = 0$ by stopping inflow into the mainline altogether until $x(T)\in\mc X_f$ in finite time. Under $u\equiv0$, when $\bar{x}^{\text{jam}} = \ubar{x}^{\text{jam}}$, the only equilibrium of $F$ in \eqref{eq:mixed-monotone-decomposition-function} for the mainline state is $x_i=0,z_i=0,i\in[I]$. Consider the initial condition $x_i(0) = \bar{x}^{\text{jam}}_i, z_i(0) = 0, i\in[I]$. Then, for all $i\in[I]$, $x_i(1) \leq x_i(0)$ since $x_i(0)$ is on the boundary of $\mc X(\theta)$, and $z_i(1) = z_i(0) = 0$. By the monotonicity property in Proposition~\ref{prop:mixed-monotone}, $x_i(k+1) \leq x_i(k)$ for all $k\geq0$ and the trajectory will converge to the equilibrium zero. Therefore, the mainline trajectory starting from any initial condition such that $x_i(0)\leq \bar{x}^{\text{jam}}_i, z_i(0)\geq 0, x_i(0) \geq z_i(0),i\in[I]$ will converge to zero under $\hat{u}(k)=0$ for all $k\geq0$.

For case \ref{prop:feasible-control-non-empty-condition-2}, it is sufficient to provide proof for the special case when $\xter_i=x^{\text{up}}_i,i\in[I]$ and $\xter_i=0,i\in[I+1\colon 2I]$. Let $\lambda=\lambdaup = \lambdalb$, we write $\xunc(\lambda)$ as $\xunc$ for simplicity.
 Consider the following control policy: 
 \begin{equation}\label{eq:vanishing-queue-policy}
 	\begin{aligned}
 		&\text{Step }1\colon  \hat{u}_i(k) =0, \quad \text{if }\exists \, j \in [I] \text{ s.t. } x^m_j(k)>\xunc_j \\
 		&\text{Step }2\colon  \hat{u}_i(k)  \\
 		&=\begin{cases}
 			\min\{x_{i}^r(k)+\lambda_i, v_i \, \xunc_i\},\ \\ \hspace{0.1in}  i=1,  \&\  \exists \ j\in[i+1\colon I] \text{ s.t. } x_j^r(k) >0, \\
 				\min\{x_{i}^r(k)+\lambda_i, v_i \, x^{\text{up}}_i\},\ \\ \hspace{0.1in} i=1,  \&\  x_j^r(k) = 0, j\in[i+1\colon I]\\
 			\min\{x_{i}^r(k)+\lambda_i, v_i\xunc_i -\beta_{i-1}v_{i-1}x^m_{i-1}(k)\}, \\  \hspace{0.1in} i \in [2\colon I-1],  \&\  \exists\  j\in[i+1\colon I] \text{ s.t. }  x_j^r(k) >0,\\
 			\min\{x_{i}^r(k)+\lambda_i, v_ix^{\text{up}}_i -\beta_{i-1}v_{i-1}x^m_{i-1}(k)\}, \\  \hspace{0.1in} i=I \text{ or } i \in [2\colon I-1] \ \&\  x^r_j(k) = 0, j\in[i+1\colon I] \\
 		\end{cases}
 	\end{aligned}
 \end{equation} 
 
In Step $1$, this control policy steers the mainline state $x^m_i$ to $[0, \xunc_i]$ for all $i\in[I]$ in finite time. By Proposition~\ref{prop:mixed-monotone}, after finite time steps $K$, $z_i(K)\leq x_i(K)\leq\xunc_i,i\in[I]$, and hence dynamics $F$ reduce to \eqref{eq:cell_compact_dynamic} with $\thetaup = \thetalb=\theta$. Therefore, in Step $2$, it suffices to consider one state $x$, and without loss of generality, one can assume that the initial condition satisfies $0\leq x^m\leq \xunc$.

 From an initial condition such that $0\leq x^m\leq \xunc$, we now show that, under the control policy in \eqref{eq:vanishing-queue-policy}, $[0,x^{\text{up}}_1]\times\ldots\times[0,x^{\text{up}}_I]\times[0,+\infty)^I$ is positively invariant. From $\xunc\leq\ x^{\text{up}}, x^m(0)\leq x^{\text{up}}$.
 For $i=1$, \eqref{eq:vanishing-queue-policy} implies that $\hat{u}_1(0) \leq v_1 x^{\text{up}}_1$. Therefore, $x^{\text{up}}_1 \geq x^{\text{up}}_1 - v_1x^{\text{up}}_1  + \hat{u}_1(0)\geq x^m_{1}(0)-v_1x^m_{1}(0) + \hat{u}_1(0) =  x_1^m(1)$ as $v_1\leq1$ (cf. Remark~\ref{remark:CFL}) and $x^m_1(0)\leq x^{\text{up}}_1$. For $i \geq 2$, \eqref{eq:vanishing-queue-policy} implies $\hat{u}_i(0) \leq v_ix^{\text{up}}_i -\beta_{i-1}v_{i-1}x^m_{i-1}(0)$. Therefore, $x^{\text{up}}_i \geq x^{\text{up}}_i + \beta_{i-1}v_{i-1} x^m_{i-1}(0) - v_{i}x^{\text{up}}_i+ \hat{u}_{i}(0) \geq x^m_{i}(0) + \beta_{i-1}v_{i-1} x^m_{i-1}(0) - v_{i}x^m_{i}(0) + \hat{u}_{i}(0)=x^m_{i}(1)$. The last inequality again follows from $v_i\leq1$ (cf. Remark~\ref{remark:CFL}) and $x^m_i(0)\leq x^{\text{up}}_i$. Following a similar analysis, for cell $i$ such that $ \exists\ j\in[i+1\colon I] \text{ s.t. } x^r_j(0) >0$, the control policy \eqref{eq:vanishing-queue-policy} is such that $x_i^m(1)\leq \xunc_i$ since $x_i^m(0)\leq \xunc_i$.

We now show that on-ramp queue lengths are steered to zero in finite time and remain zero afterward. 
At every time $k$, when $x^r(k)\neq0$, there exists a unique cell $i\in[I]$ such that $x_i^r(k)>0$, and $x_j^r(k) = 0,\forall\ j\in[i + 1\colon I]$ if $i\neq I$. For such a cell $i$, the on-ramp queue length is positive for all time from $0$ to current time $k$ since otherwise \eqref{eq:vanishing-queue-policy} would apply $\hat{u}_i=\lambda_i$ (for reasons specified later) after some time $\tilde{k}$ between $0$ and $k$ when $x_i^r(\tilde{k})=0$, and the queue length remains zero after $\tilde{k}$. 
We first consider the case $i\in[2\colon I]$ where $\hat{u}_i(k) = v_ix_i^{\text{up}} - \beta_{i-1}v_{i-1} x^m_{i-1}(k)$. The case where $i=1$ and $\hat{u}_1(k) = v_1x_1^{\text{up}}$ follows similarly. We first note that $x^m_{i-1}(k)\leq\xunc_{i-1}$ since $x_i^r(k) >0$ and \eqref{eq:vanishing-queue-policy} ensures $x_{i-1}^m(k)\leq\xunc_{i-1}$ if $\exists\ j\in[i+1\colon I] \text{ s.t. } x^r_j(k) >0$ by the positively invariant analysis above. Now for this cell $i$
\begin{equation*}
	\begin{aligned}
		&\quad\ x_i^r(k+1) - x_i^r(k)\\ &= \lambda_i - \hat{u}_i(k) \\ & = v_i(\xunc_i - x_i^{\text{up}} ) +  \beta_{i-1}v_{i-1}(x^m_{i-1}(k) - \xunc_{i-1})
		\\ & \leq v_i(\xunc_i - x_i^{\text{up}} ) <0
	\end{aligned}
\end{equation*} 
where the second equality follows from the definition of $\xunc$ and the last inequality follows from the definition of $x^{\text{up}}$. The queue length strictly decreases at a constant rate until $k=\tilde{K}$ such that $\hat{u}_i(\tilde{K}) = x_i^r(\tilde{K})+\lambda_i$. Then $x_i^r(\tilde{K}+1)=x_i^r(\tilde{K})+\lambda_i - \hat{u}_i(\tilde{K})=x_i^r(\tilde{K})+\lambda_i - (x_i^r(\tilde{K})+\lambda_i) = 0$, i.e., the queue length reaches zero in at most one time step and $x_i^r(\tilde{K}+1)=0$. 
 
 Lastly, we show that $\hat{u}_i(\tilde{K}+1) = \lambda_i$ and thus the queue length remains zero. For cell $i=1$, this follows from $x^{\text{up}}_1\geq\xunc_1$.  For cell $i\in[2\colon I]$, it suffices to show that $v_ix_i^{\text{up}} - \beta_{i-1}v_{i-1} x^m_{i-1}\geq \lambda_i, \forall\, x^m_{i-1}\in[0,x^{\text{up}}_{i-1}]$. This follows by $\beta_{i-1}v_{i-1}x^m_{i-1}\leq\beta_{i-1}v_{i-1}x^{\text{up}}_{i-1}\leq v_{i}(x^{\text{up}}_{i} - \xunc_{i}) + \beta_{i-1}v_{i-1}\xunc_{i-1}$ from the definition of $x^{\text{up}}$, and $\lambda_i = v_i\xunc_i - \beta_{i-1}v_{i-1}\xunc_{i-1}$. 
 Therefore, the queue length of every on-ramp vanishes after finite time steps and remains zero under $\hat{u}(k)=\lambda$.

\subsection{Proof of Proposition~\ref{prop:recursive-feasibility}}
\label{proof:recursive-feasibility}
	We proceed by showing that $\mathbbm{P}(\hat{\xlift}(t),\lambdalift(t), \thetalift(t))$ is feasible $\implies$ $\mathbbm{P}(\tilde{\xlift}(t+1), \lambdalift(t), \thetalift(t))$ is feasible $\implies$ $\mathbbm{P}(\hat{\xlift}(t + 1), \lambdalift(t+1), \thetalift(t+1))$ is feasible for any $t\geq0$.  Let $\{\hat{u}^*(0\colon T-1| t)\}$ denote an optimal solution to $\mathbbm{P}(\hat{\xlift}(t),\lambdalift(t),  \thetalift(t))$ and let $\tilde{u}(T|t)$ be a control such that $\Xflift$ is positively invariant.
	Let $\begin{bmatrix} \Phi_{\bar{x}}(\xlift_\circ,u(0\colon k-1),\lambdalift;\thetalift) \\ \Phi_{\ubar{x}}(\xlift_\circ,u(0\colon k-1),\lambdalift;\thetalift) \end{bmatrix}=: \pmb{\Phi}(\xlift_\circ,u(0\colon k-1),\lambdalift;\thetalift)$ denote the state transition function of dynamics $\Flift$ in \eqref{eq:mixed-monotone-decomposition-function} at time $k$ with initial condition $\xlift_\circ$, control inputs $u(0\colon k-1)$, demand $\lambdalift$, and parameters $\thetalift$. 
	Since state space $\Xlift(\thetalift)$ only depends on jam density and jam density is known by Assumption~\ref{assu:initial-condition}, we omit the dependence of $\Xlift(\thetalift)$ on $\thetalift$.
	
	We first show that the sequence $\{\hat{u}^*(1\colon T-1| t),\tilde{u}(T|t)\}$ is  a feasible solution to $\mathbbm{P}(\tilde{\xlift}(t+1), \lambdalift(t), \thetalift(t))$.
	 From \eqref{eq:total-framework}, we have
	 $\pmb{\Phi}(\hat{\xlift}(t), \hat{u}^*(0|t),\lambdalift(t); \thetalift(t)) = \tilde{\xlift}(t+1).$
	 Therefore,
	 $\pmb{\Phi}(\hat{\xlift}(t), \hat{u}^*(0\colon k - 1| t),\lambdalift(t); \thetalift(t)) =  \pmb{\Phi}(\tilde{\xlift}(t+1), \hat{u}^*(1\colon k-1| t),\lambdalift(t); \thetalift(t))\in\Xlift$ for all $k\in[2:T]$. Since $\pmb{\Phi}(\hat{\xlift}(t), \hat{u}^*(0\colon T-1| t),\lambdalift(t); \thetalift(t))\in\Xflift$ and $\Xflift$ is positively invariant under the control $\tilde{u}(T|t)$, we have $\pmb{\Phi}(\tilde{\xlift}(t+1), \{\hat{u}^*(0\colon T-1| t),\tilde{u}(T|t)\},\lambdalift(t); \thetalift(t))\in\Xflift$. 
	
	Then, we show that $\{\hat{u}^*(1\colon T-1| t),\tilde{u}(T|t)\}$ is also a feasible solution to $\mathbbm{P}(\hat{\xlift}(t + 1), \lambdalift(t+1),\thetalift(t+1))$. By Lemma~\ref{lemm:set-membership} and Proposition~\ref{prop:mixed-monotone}, we have $\bar{\theta}(t+1) \leq \bar{\theta}(t), \ubar{\theta}(t+1)\geq\ubar{\theta}(t),\bar{\lambda}(t+1) \leq \bar{\lambda}(t), \ubar{\lambda}(t+1)\geq\ubar{\lambda}(t)$, and $\xup(t+1)\leq\tilde{\bar{x}}(t+1), \xlb(t+1) \geq \tilde{\bar{x}}(t+1)$.  Therefore, $ \Phi_{\bar{x}}(\hat{\xlift}(t+1), \hat{u}^*(1\colon k-1| t),\lambdalift(t+1); \thetalift(t+1))\leq\Phi_{\bar{x}}(\tilde{\xlift}(t+1), \hat{u}^*(1\colon k-1| t),\lambdalift(t); \thetalift(t))\in\mc X$ and thus $\Phi_{\bar{x}}(\hat{\xlift}(t+1), \hat{u}^*(1\colon k-1| t),\lambdalift(t+1); \thetalift(t+1))\in\mc X$ for all $k\in[2:T]$ since $\mc X$ is box-constrained. Also, $\Phi_{\bar{x}}(\tilde{\xlift}(t+1), \{\hat{u}^*(1\colon T-1| t),\tilde{u}(T|t)\},\lambdalift(t); \thetalift(t))\in\mc X_f$ implies that $\Phi_{\bar{x}}(\hat{\xlift}(t+1), \{\hat{u}^*(1\colon T-1| t),\tilde{u}(T|t)\},\lambdaup(t+1); \thetalift(t+1))\in\mc X_f$ since $\mc X_f$ is also box-constrained.
\subsection{Proof of Theorem~\ref{theo:stability-boundedness}}
	\label{proof:boundeness-stability}
	We first show that the closed-loop system trajectory enters the terminal set in finite time.
	For any feasible solution $\hat{u}(0: T-1| t)$, let $V(\hat{\xlift}(t), \hat{u}(0: T-1| t),\lambdalift(t), \thetalift(t))$ denote the objective value of \eqref{eq:mpc-opt-adaptive}.
	Let $V^*(\hat{\xlift}(t),\lambdalift(t), \thetalift(t))$ denote the optimal value function of \eqref{eq:mpc-opt-adaptive}. Now we show that the value function $V^*(\hat{\xlift}(t),\lambdalift(t),\thetalift(t))$ satisfies $V^*(\hat{\xlift}(t + 1),\lambdalift(t+1), \thetalift(t+1)) - V^*(\hat{\xlift}(t),\lambdalift(t),\thetalift(t)) \leq -\ell(\xup(t),u(t))$ so that the optimal objective value of \eqref{eq:mpc-opt-adaptive} decreases by at least $\ell(\xup(t),u(t))$ at each time step $t$. Let $\hat{u}^*(0: T-1| t)$ be an optimal solution to $\mathbbm{P}(\hat{\xlift}(t),\lambdalift(t),\thetalift(t))$ and $\tilde{u}(T) = 0$. 
	Let $\hat{u}(0: T-1|t+1)\equiv\{\hat{u}^*(1: T-1| t),\tilde{u}(T)\}$ denote a sequence of controls where $\hat{u}(0: k|t+1) = \hat{u}^*(1: k+1| t), k\in[0:T-2]$ and $\hat{u}(T-1|t+1) = \tilde{u}(T)$. 
	The sequence $\hat{u}(0: T-1|t+1)$ is also a feasible solution to $\mathbbm{P}(\hat{\xlift}(t+1),\lambdalift(t+1),\thetalift(t+1))$ by the proof of Proposition~\ref{prop:recursive-feasibility} and Example~\ref{ex:positively-invariant}. 
	Also, the cost function is monotone in state and  $\Phi_{\bar{x}}(\hat{\xlift}(t+1),  \hat{u}(0: k-1| t+1),\lambdalift(t+1); \thetalift(t+1)) \leq \Phi_{\bar{x}}(\tilde{\xlift}(t+1), \hat{u}(0: k-1| t+1),\lambdalift(t); \thetalift(t)),\forall\, k\in[1: T]$\footnote{The notation is defined in the Proof of Proposition~\ref{prop:feasible-control-non-empty}.}. Therefore, we have:
	\begin{equation}\label{eq:value-function-t-adaptive}
		\begin{split}
			&\quad\ V^*(\hat{\xlift}(t+1),\lambdalift(t+1), \thetalift(t+1))\\&\leq V(\hat{\xlift}(t+1), \hat{u}(0: T-1| t+1),\lambdalift(t+1), \thetalift(t+1)) \\ 
			& \leq V(\tilde{\xlift}(t+1), \hat{u}(0: T-1| t+1),\lambdalift(t),\thetalift(t)) \\
		\end{split}
	\end{equation}
	This implies
	\begin{equation*}
		\begin{split}
			&V^*(\hat{\xlift}(t+1),\lambdalift(t+1), \thetalift(t+1)) - V^*(\hat{\xlift}(t),\lambdalift(t), \thetalift(t)) \\
			\leq\, & V(\tilde{\xlift}(t+1), \hat{u}(0: T-1| t+1),\lambdalift(t), \thetalift(t)) \\ &\hspace{1in} -V(\hat{\xlift}(t), \hat{u}^*(0: T-1| t),\lambdalift(t), \thetalift(t)) \quad 
	   \end{split}
    \end{equation*}
    \begin{equation}\label{eq:value-function-inequality}
        \begin{split}
			=\, & V_f(\Phi_{\bar{x}}(\tilde{\xlift}(t+1), \hat{u}(0: T-1| t+1),\lambdalift(t);\thetalift(t)) \\
			& - V_f(\Phi_{\bar{x}}(\hat{\xlift}(t), \hat{u}^*(0: T-1| t),\lambdalift(t);\thetalift(t))) \\
			& + \ell\big(\Phi_{\bar{x}}(\tilde{\xlift}(t+1), \hat{u}(0: T-2| t+1),\lambdalift(t);\thetalift(t)),\\ & \hspace{2in} \hat{u}(T-1| t+1)\big) \\
			& +\sum_{k=1}^{T-2} \Big[\ell\big(\Phi_{\bar{x}}(\tilde{\xlift}(t+1), \hat{u}(0: k-1| t+1),\lambdalift(t);\thetalift(t)),\\ & \hspace{2in} \hat{u}(k| t+1)\big) \\ 
			&\hspace{0.2in} - \ell(\Phi_{\bar{x}}(\hat{\xlift}(t), \hat{u}^*(0: k| t),\lambdalift(t);\thetalift(t)), \hat{u}^*(k+1| t))\Big] \\
			& + \ell(\tilde{\bar{x}}(t+1), \hat{u}(0| t+1)) \\ & \hspace{0.5in} - \ell(\Phi_{\bar{x}}(\hat{\xlift}(t), \hat{u}^*(0| t),\lambdalift(t);\thetalift(t)), \hat{u}^*(1| t))\\ & - \ell(\xup(t), \hat{u}^*(0|t)) \\
            =\, & V_f(\Phi_{\bar{x}}(\tilde{\xlift}(t+1), \hat{u}(0: T-1| t+1),\lambdalift(t);\thetalift(t)) \\
			& - V_f(\Phi_{\bar{x}}(\hat{\xlift}(t), \hat{u}^*(0: T-1| t),\lambdalift(t);\thetalift(t))) \\
			& + \ell\big(\Phi_{\bar{x}}(\tilde{\xlift}(t+1), \hat{u}(0: T-2| t+1),\lambdalift(t);\thetalift(t)),\\ & \hspace{2in} \hat{u}(T-1| t+1)\big) \\
            & - \ell(\xup(t), \hat{u}^*(0|t)) \\
		\end{split}
	\end{equation} 
	where the last equality follows from 
    $\pmb{\Phi}(\hat{\xlift}(t), \hat{u}^*(0: k - 1| t),\lambdalift(t); \thetalift(t)) =  \pmb{\Phi}(\tilde{\xlift}(t+1), \hat{u}^*(1: k-1| t),\lambdalift(t); \thetalift(t))\in\Xlift,k\in[2:T]$.

    By the assumptions in Theorem~\ref{theo:stability-boundedness}, $\ell(x,u)=0$ if $x\in\mc X_f$ and $V_f\equiv0$. Since the terminal constraint implies $\Phi_{\bar{x}}(\tilde{\xlift}(t+1), \hat{u}(0: T-2| t+1),\lambdalift(t);\thetalift(t))=\Phi_{\bar{x}}(\hat{\xlift}(t),\hat{u}^*(0: T-1| t),\lambdalift(t);\thetalift(t))\in\mc X_f$, we have
	\begin{multline}\label{eq:value-function-lyapunov}
		V^*(\hat{\xlift}(t+1),\lambdalift(t+1), \thetalift(t+1)) - V^*(\hat{\xlift}(t),\lambdalift(t), \thetalift(t))\\ \leq -\ell(\xup(t), \hat{u}^*(0|t)) = -\ell(\xup(t),u(t))
	\end{multline}
	Suppose $x(0)\notin\mc X_f$, which implies $\xup(0)\notin\mc X_f$. Then, we show that there exists a finite time $K$ such that $\xup(K)\in\mc X_f$, which implies $x(K)\in\mc X_f$. Let $l_{\min} := \min_i\{l_i:i\in[I]\}, x_{\min}^f := \min_i\{\xter_i:i\in[I]\}$. For $x\notin\mc X_f$, we have $\ell(x,u) = l^T x\geq l_{\min}x_{\min}I =: \ell_{\min}^f>0$ since $l$ and $\xter$ are positive, and $\ell(x,u)= 0$ for $x\in\mc X_f$. Let $\bar{t}$ denote a sufficiently large but finite integer such that $\bar{t}\ell_{\min}^f> V^*(\hat{\xlift}(0),\lambdalift(0), \thetalift(0))$. Suppose the state has not entered $\mc X_f$ by time $t=\bar{t}$, we have $\ell(\xup(t),u(t))\geq \ell_{\min}^f$ for all $t\in[0:\bar{t}]$. From the telescoping sum of the inequality \eqref{eq:value-function-inequality} from $t=0$ to $\bar{t}-1$, we have $V^*(\hat{\xlift}(\bar{t}),\lambdalift(\bar{t}), \thetalift(\bar{t})) - V^*(\hat{\xlift}(0),\lambdalift(0), \thetalift(0))\leq -\sum_{t=0}^{\bar{t}}\ell(\xup(t),u(t)) \leq -\bar{t}\ell_{\min}^f$. This implies that $V^*(\hat{\xlift}(\bar{t}),\lambdalift(\bar{t}), \thetalift(\bar{t}))\leq V^*(\hat{\xlift}(0),\lambdalift(0), \thetalift(0)) -\bar{t}\ell_{\min}^f <0 $, which contradicts the fact that $V^*(\hat{\xlift}(\bar{t}), \lambdalift(\bar{t}), \thetalift(\bar{t}))$ is nonnegative. Therefore, $x(K)\in\mc X_f$ for some time $K\leq\bar{t}$.
	
	The proof for the second part of Theorem~\ref{theo:stability-boundedness} is as follows.
	Let $\xter_i\leq\xunc_i(\lambda),i\in[I]$. When $x(t)\in\mc X_f$ and thus $x_i(t)\leq \xunc_i(\lambda),i\in[I]$, we have $\fout_{I+i}(x(t),u(t)) = u_i(t),i\in[I]$ and the dynamics reduce to  \eqref{eq:cell_compact_dynamic}.
	To show that $u(t)=y^r(t) - y^r(t-1) + u(t-1)$ stabilizes the system starting with initial conditions in $\mc X_f$, we first observe that $\lambda = y^r(t) - y^r(t-1) + u(t-1)$ from dynamics \eqref{eq:cell_compact_dynamic}. Therefore, the control becomes $u(t)=\lambda$ and $x^r(t+1) = x^r(t)$ under such control. Also, the set of mainline states $\{x^m: 0\leq x^m\leq \xunc(\lambda)\}$ is positively invariant for dynamics \eqref{eq:cell_compact_dynamic} under $u(t) = \lambda$. This follows by $x^m_i(t+1) = f_i(x^m(t),x^r(t),u(t),\lambda;\theta)\leq f_i(\xunc(\lambda),x^r(t),u(t),\lambda;\theta) = \xunc_i(\lambda),i\in[I]$ for all $x^m(t)\leq\xunc(\lambda)$, which is due to the monotonicity of  \eqref{eq:cell_compact_dynamic} by Lemma~\ref{lemm:monotone_x} and the fact that $\xunc(\lambda)$ is an equilibrium for the mainline state under $u(t)=\lambda$.
	 Therefore, $||x(t)||<\infty$.

\subsection{Proof of Theorem~\ref{theo:ISpS}}
\label{sec:ISS-stability-proof}
We omit the dependence of value function $V^*$
\footnote{The notation is defined in the Proof of Theorem~\ref{theo:stability-boundedness}.} 
and state transition function $\Phi_{\bar{x}}$
\footnote{The notation is defined in the Proof of Proposition~\ref{prop:feasible-control-non-empty}.} 
on $\thetalift$ and $\lambdalift$ since they are assumed to be known.
Let $\tilde{u}(T)$ be a control such that \eqref{eq:terminal-lyapunov} holds.
By the assumptions in Theorem~\ref{theo:ISpS}, $V_f(x) = b^Tx$ and $\ell(x,u)=l^Tx$. Then, following the Proof of Theorem~\ref{theo:stability-boundedness}, 
\begin{equation}\label{eq:value-function-inequality-theorem2}
	\begin{aligned}
		&V^*(\hat{\xlift}(t+1)) - V^*(\hat{\xlift}(t)) \\
	\leq \ & b^T\Phi_{\bar{x}}(\hat{\xlift}(t),\{\hat{u}^*(0:T-1| t),\tilde{u}(T)\} )  \\ 
	 & - (b-l)^T \Phi_{\bar{x}}(\hat{\xlift}(t),\hat{u}^*(0:T-1| t))	- l^T\xup(t) \quad \text{by \eqref{eq:value-function-inequality}}\\
	\leq & -l^T\xup(t) + d^T\lambda 
	\end{aligned}
\end{equation} 
where the last inequality follows by $\Phi_{\bar{x}}(\hat{\xlift}(t),\hat{u}^*(0\colon T-1| t))\in\mc X_f$ and \eqref{eq:terminal-lyapunov}.
We derive the following bounds on the value function:
 \begin{equation}\label{eq:value-function-comparison}
     \tilde{a}_1\lVert \xup(t)\rVert_1\leq V^*(\hat{\xlift}(t)) \leq \tilde{a}_2 \lVert \xup(t)\rVert_1 +\tilde{a}_2\lVert \lambda\rVert_1
 \end{equation} 
 where $\tilde{a}_1 = \min_i\{l_i\}$, 
$\tilde{a}_2 =  (T+1)\max\{ \max_i \,l_i , \max_i \, b_i\}$.
The lower bound follows from $\tilde{a}_1\lVert \xup(t)\rVert_1\leq l^T\xup(t)$. The upper bound follows from $||\xup(t+1)||_1\leq||\xup(t)||_1 + ||\lambda||_1$.
From \eqref{eq:value-function-inequality-theorem2}, $V^*(\hat{\xlift}(t+1))-V^*(\hat{\xlift}(t))\leq -\tilde{a}_1\lVert \xup(t)\rVert_1 + \tilde{a}_3\lVert  \lambda \rVert_1$ holds with 
$\tilde{a}_3 = \max_i\{d_i\}$. 
Then we have 
\begin{align*}
    &V^*(\hat{\xlift}(t+1))\\ &\leq V^*(\hat{\xlift}(t)) - \tilde{a}_1\lVert \xup(t)\rVert_1 + \tilde{a}_3\lVert \lambda \rVert_1
    \\ & \leq V^*(\hat{\xlift}(t)) -\dfrac{\tilde{a}_1\lVert \xup(t)\rVert_1}{\tilde{a}_2\lVert \xup(t) \rVert_1} (V^*(\hat{\xlift}(t)) - \tilde{a}_2\lVert \lambda\rVert_1) + \tilde{a}_3\lVert \lambda \rVert_1 \\ &\hspace{0.5in} \text{ by \eqref{eq:value-function-comparison} that implies }\frac{(V^*(\hat{\xlift}(t)) - \tilde{a}_2\lVert \lambda\rVert_1)}{\tilde{a}_2\lVert \xup(t) \rVert_1}\leq1 \\
    &= \rho V^*(\hat{\xlift}(t)) + (1-\rho)\tilde{a}_2\lVert \lambda\rVert_1 + \tilde{a}_3\lVert \lambda \rVert_1
\end{align*}
 where $\rho:=1-\tilde{a}_1/\tilde{a}_2$ and $\rho\in(0,1)$. Tracing back to time $0$, we have the following:
\begin{equation*}
	\begin{aligned}
		 &V^*(\hat{\xlift}(t+1)) \leq \rho^{t+1}V^*(\hat{\xlift}(0)) +\\ & \hspace{1in}  \sum_{k=0}^{t}\rho^k(\tilde{a}_3 + (1-\rho)\tilde{a}_2)\lVert \lambda\rVert_1,\quad \forall t\geq 0
	\end{aligned}
\end{equation*}
By \eqref{eq:value-function-comparison} and $\rho\in(0,1)$, we have 
\begin{equation*}
	\begin{aligned}
		     &\lVert \xup(t+1)\rVert_1 \leq \dfrac{\tilde{a}_2}{\tilde{a}_1} \rho^{t+1}(\lVert \xup(0)\rVert_1+ \lVert \lambda \rVert_1)+\\ 
		     & \hspace{1in}  \dfrac{(\tilde{a}_3+(1-\rho)\tilde{a}_2)(1-\rho^{t+1})}{\tilde{a}_1(1-\rho)}\lVert \lambda \rVert_1, \\
		     &\hspace{0.45in} \leq \dfrac{\tilde{a}_2}{\tilde{a}_1} \rho^{t+1}\lVert \xup(0)\rVert_1 + \dfrac{\tilde{a}_3 + (1-\rho)\tilde{a}_2}{\tilde{a}_1(1-\rho)}\lVert \lambda \rVert_1, \forall t\geq 0
	\end{aligned}
\end{equation*}
Then, \eqref{eq:ISS-bound} follows from the above inequality and $x(t)\leq\xup(t),\xup(0)\leq\bar{x}(0)$.

\end{document}